\documentclass[journal]{IEEEtran}

\usepackage{graphicx} % support the \includegraphics command and options
\usepackage{array} % for better arrays (eg matrices) in maths
\usepackage{multirow}
\usepackage{subfigure} 
\usepackage{float}
\usepackage{amsmath}
\usepackage{amsthm}
\usepackage{amssymb}
\usepackage{threeparttable}
\usepackage{bbold}
\usepackage{balance}

\usepackage{upquote}
\usepackage{url, listings}

\usepackage[linesnumbered,ruled,vlined]{algorithm2e}
\SetKwProg{Fn}{Function}{}{}

\newcommand{\nop}[1]{}
\usepackage{xcolor}

\begin{document}

\title{A Survey on Edge Computing Systems and Tools}

\author{
		Fang~Liu,
		Guoming~Tang,
		Youhuizi~Li,
		Zhiping~Cai,
		Xingzhou~Zhang,
		Tongqing~Zhou
\thanks{F. Liu is with the School of Data and Computer Science, Sun Yat-sen University, Guangzhou, Guangdong, China (e-mail: {liufang25}@mail.sysu.edu.cn).
G. Tang is with the Key Laboratory of Science and Technology on Information System Engineering, National University of Defense Technology, Changsha, Hunan, China (e-mail: {gmtang}@nudt.edu.cn).
Y. Li is with the School of Compute Science and Technology, Hangzhou Dianzi University, China.
Z. Cai and T. Zhou are with the College of Computer, National University of Defense Technology, Changsha, Hunan, China (e-mail: {zpcai}@nudt.edu.cn, {zhoutongqing1991}@163.com).
X. Zhang is with State Key Laboratory of Computer Architecture, Institute of Computing Technology, Chinese Academy of Sciences, China.  
}% <-this % stops a space
%\thanks{}
}

\maketitle

\begin{abstract}
Driven by the visions of Internet of Things and 5G communications, the edge computing systems integrate computing, storage and network resources at the edge of the network to provide computing infrastructure, enabling developers to quickly develop and deploy edge applications. Nowadays the edge computing systems have received widespread attention in both industry and academia. To explore new research opportunities and assist users in selecting suitable edge computing systems for specific applications, this survey paper provides a comprehensive overview of the existing edge computing systems and introduces representative projects. A comparison of open source tools is presented according to their applicability. Finally, we highlight energy efficiency and deep learning optimization of edge computing systems. Open issues for analyzing and designing an edge computing system are also studied in this survey.
\end{abstract}

\section{Introduction}\label{sec:intro}

In the post-Cloud era, the proliferation of Internet of Things (IoT) and the popularization of 4G/5G, gradually changes the public's habit of accessing and processing data, and challenges the linearly increasing capability of cloud computing. Edge computing is a new computing paradigm with data processed at the edge of the network. Promoted by the fast growing demand and interest in this area, the edge computing systems and tools are blooming, even though some of them may not be popularly used right now. 

\begin{figure}[!tb]
\begin{center}
\includegraphics[width=0.42\textwidth]{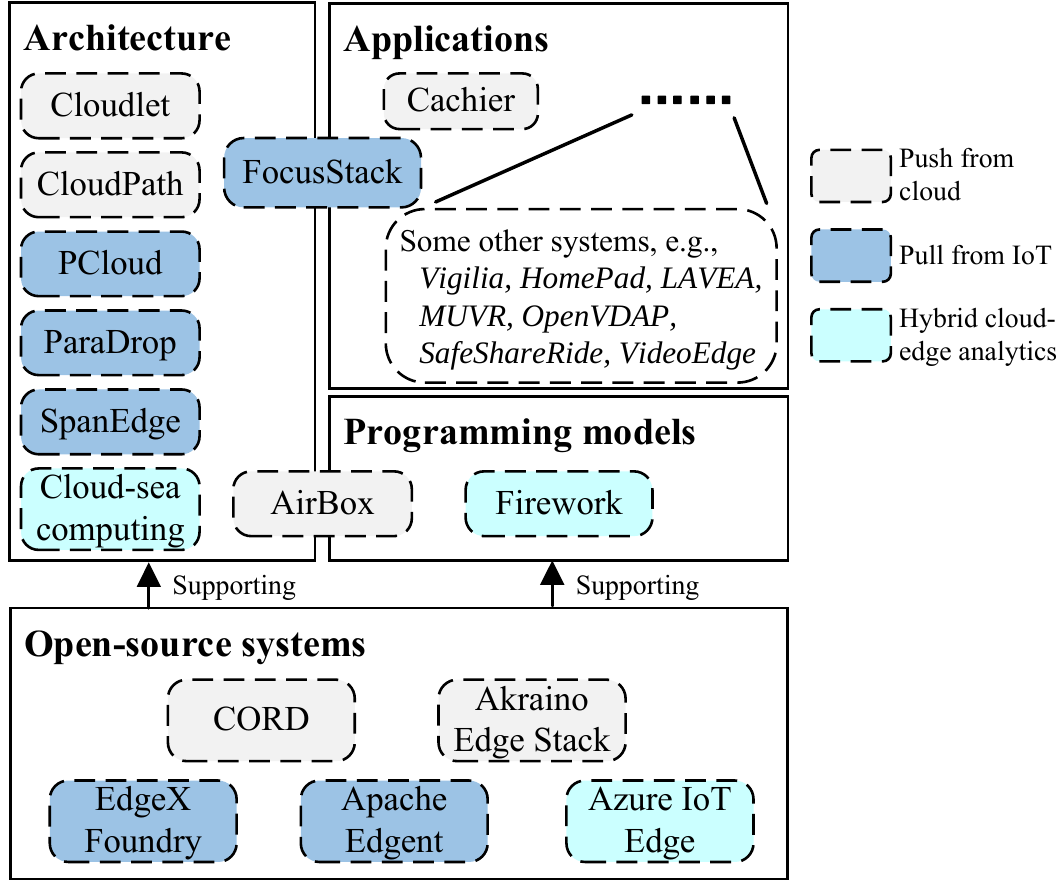}
\caption{Categorization of edge computing systems.}\label{fig:systemCategory}
%\vspace{-0.20in}
\end{center}
\end{figure}

There are many classification perspectives to distinguish different edge computing system. To figure out why edge computing appears as well as its necessity, we pay more attention to the basic motivations. Specifically, based on different design demands, existing edge computing systems can roughly be classified into three categories, together yielding innovations on system architecture, programming models and various applications, as shown in Fig.~\ref{fig:systemCategory}. %\nop{Specifically, the three categories are : 1) push from cloud. Cloud providers push services and computation to the edge in order to reduce response time, improve user experience and leverage locality better; 2) pull from IoT. The IoT \textcolor{red}{devices} at the edge generate a huge amount of data which is impossible to be transmitted to the cloud. Due to the real-time response requirement of modern IoT applications, IoT pull the services and computation from the faraway cloud to the near edge to better support their usage; 3) hybrid cloud-edge analytics. Modern advanced services and applications require both global optimal results and minimum response time, which can only be achieved by combining the advantages of cloud and edge. Fig.~\ref{fig:frameworkIntro} shows representative systems in each category.}  

%category.jpg

\begin{itemize}
\item Push from cloud. In this category, cloud providers push services and computation to the edge in order to leverage locality, reduce response time and improve user experience. Representative systems include Cloudlet, Cachier, AirBox, and CloudPath. Many traditional cloud computing service providers are actively pushing cloud services closer to users, shortening the distance between customers and cloud computing, so as not to lose market to mobile edge computing. For example, Microsoft launched AzureStack in 2017, which allows cloud computing capabilities to be integrated into the terminal, and data can be processed and analyzed on the terminal device.
\item Pull from IoT. Internet of Things (IoT) applications pull services and computation from the faraway cloud to the near edge to handle the huge amount of data generated by IoT devices. Representative systems include PCloud, ParaDrop, FocusStack and SpanEdge. Advances in embedded Systems-on-a-Chip (SoCs) have given rise to many IoT devices that are powerful enough to run embedded operating systems and complex algorithms. Many manufacturers integrate machine learning and even deep learning capabilities into IoT devices. Utilizing edge computing systems and tools, IoT devices can effectively share computing, storage, and network resources while maintaining a certain degree of independence.
\item Hybrid cloud-edge analytics. The integration of advantages of cloud and edge provides a solution to facilitate both global optimal results and minimum response time in modern advanced services and applications. Representative systems include Firework and Cloud-Sea Computing Systems\nop{ and AWS Greengrass}. Such edge computing systems utilize the processing power of IoT devices to filter, pre-process, and aggregate IoT data, while employing the power and flexibility of cloud services to run complex analytics on those data. For example, Alibaba Cloud launched its first IoT edge computing product, LinkEdge, in 2018, which expands its advantages in cloud computing, big data and artificial intelligence to the edge to build a cloud/edge integrated collaborative computing system; Amazon released AWS Greengrass in 2017, which can extend AWS seamlessly to devices so that devices can perform local operations on the data they generate, while data is transferred to the cloud for management, analysis, and storage.
\end{itemize}
From a research point of view, this paper gives a detailed introduction to the distinctive ideas and model abstractions of the aforementioned edge computing systems. Noted that the three categories are presented to clearly explain the necessity of edge computing, and the classification is not the main line of this paper. Specifically, we review systems designed for architecture innovation first, then introduce those for programming models and applications (in Sec.~\ref{sec:edge-computing-systems}). Besides, some recently efforts for specific application scenarios are also studied.
\nop{This paper is from a research point of view, focusing on distinctive ideas and model abstractions of different systems. It introduces representative systems in the categories discussed above, and some recently proposed edge systems are outlined.}

While we can find a lot of systems using edge computing as the building block, there still lacks standardization to such a paradigm. Therefore, a comprehensive and coordinated set of foundational open-source systems/tools are also needed to accelerate the deployment of IoT and edge computing solutions. Some open source edge computing projects have been launched recently (e.g., CORD). As shown in Fig.~\ref{fig:systemCategory}, these systems can support the design of both architecture and programming models with useful APIs. We review these open source systems with a comparative study on their characteristics (in Sec.~\ref{sec:open-source-projects}).
\nop{This paper studies current work on open source edge computing tools, which is challenging and has attracted a lot of attentions.}

When designing the edge computing system described above, energy efficiency is always considered as one of the major concerns as the edge hardware is energy-restriction. Meanwhile, the increasing number of IoT devices is bringing the growth of energy-hungry services. Therefore, we also review the energy efficiency enhancing mechanisms adopted by the state-of-the-art edge computing systems from the three-layer paradigm of edge computing (in Sec.~\ref{sec:energy-efficiency}).
\nop{This paper reviews the energy efficiency enhancing mechanisms adopted by the state-of-the-art edge computing systems from three Layer of the edge computing paradigm (cloud, edge server and device).}

\begin{figure}[!tb]
\begin{center}
\includegraphics[width=0.42\textwidth]{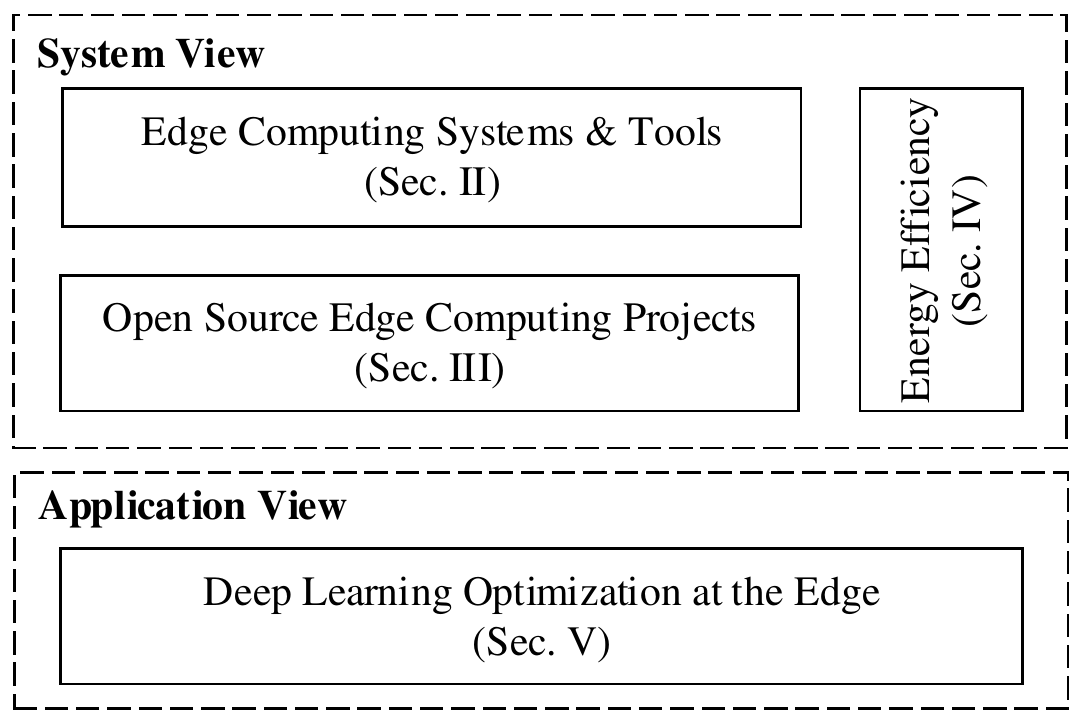}
\caption{Major building blocks and organization of this survey paper.}\label{fig:framework-intro}
\vspace{-0.20in}
\end{center}
\end{figure}

In addition to investigations from the system view, we also look into the emerging techniques in edge computing system from the application view. Recently, deep learning based Artificial Intelligence (AI) applications are widely used and offloading the AI functions from the cloud to the edge is becoming a trend. However, deep learning models are known for being large and computationally expensive. Traditionally, many systems and tools are designed to run deep learning models efficiently on the cloud. As the multi-layer structure of deep learning, it is appropriate for edge computing paradigm, and more of its functions can be offloaded to the edge. Accordingly, this paper also studies the new techniques recently proposed to support the deep learning models at the edge (in Sec.~\ref{sec:deep-learning-opt}).
\nop{This paper studies new techniques recently proposed to support the deep learning models at the edge; these techniques are categorized into three types: systems and toolkits, deep learning packages, and hardware.}

Our main contributions in this work are as follows:
\begin{itemize}
\item Reviewing existing systems and open source projects for edge computing by categorizing them from their design demands and innovations. We study the targets, architecture, characteristics, and limitations of the systems in a comparative way.
\item Investigating the energy efficiency enhancing mechanism for edge computing from the view of the cloud, the edge servers, and the battery-powered devices.
\item Studying the technological innovations dedicated to deploying deep learning models on the edge, including systems and toolkits, packages, and hardware.
\item Identifying challenges and open research issues of edge computing systems, such as mobility support, multi-user fairness, and privacy protection.
\end{itemize}

We hope this effort will inspire further research on edge computing systems. The contents and their organization in this paper are shown in Fig.~\ref{fig:framework-intro}. Besides the major four building blocks (Sec.~\ref{sec:edge-computing-systems}$\sim$Sec.~\ref{sec:deep-learning-opt}), we also give a list of open issues for analyzing and designing an edge computing system in Sec.~\ref{sec:key-design-issues}. The paper is concluded in Sec.~\ref{sec:conclusion}.

\nop{The remaining parts of this paper are organized as follows. Section II introduces ten representative edge computing systems in academia, and outline some recently proposed edge systems. Section III presents some open source edge computing projects. Section IV discusses the energy efficiency enhancing mechanisms of edge computing systems. Section V presents new techniques that support the deep learning models at the edge. Open issues for analyzing and designing an edge computing system are studied in section VI. Finally, the paper is concluded in section VII.}

\section{Edge Computing Systems and Tools}
\label{sec:edge-computing-systems}

In this section, we review edge computing systems and tools presenting architecture innovations, programming models, and applications, respectively. For each part, we introduce work under the ``push", ``pull", and ``hybrid" demand in turn.

\subsection{Cloudlet}

\begin{figure}[!tb]
\begin{center}
\includegraphics[width=0.5\textwidth]{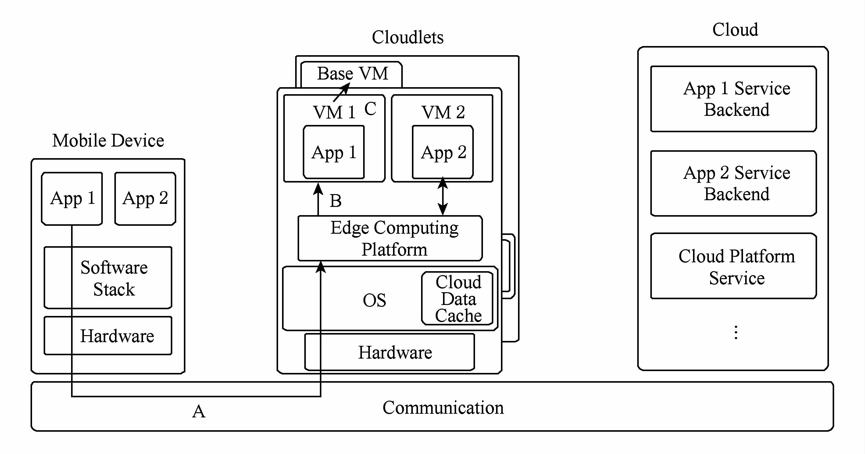}
\caption{Cloudlet Component Overview and Functions that support application mobility. 
A: Cloudlet Discovery, B: VM Provisioning, C: VM Handoff.}\label{fig:Cloudlet}
%\vspace{-0.20in}
\end{center}
\end{figure}

In 2009, Carnegie Mellon University proposed the concept of Cloudlet~\cite{satyanarayanan2009case}, and the Open Edge computing initiative was also evolved from the Cloudlet project~\cite{OpenEdgeCompute}. Cloudlet is a trusted, resource-rich computer or cluster of computers that are well-connected to the Internet and available to nearby mobile devices. It upgrades the original two-tier architecture ``Mobile Device-Cloud'' of the mobile cloud computing to a three-tier architecture ``Mobile Device-Cloudlet-Cloud''. Meanwhile, Cloudlet can also serve users like an independent cloud, making it a ``small cloud'' or ``data center in a box''. Although the Cloudlet project is not proposed and launched in the name of edge computing, its architecture and ideas fit those of the edge computing and thus can be regarded as an edge computing system. 

The Cloudlet is in the middle layer of the three-tier edge computing architecture and can be implemented on a personal computer, low-cost server, or small cluster. {It can be composed of a single machine or small clusters consisting of multiple machines.} Like WiFi service access points, a Cloudlet can be deployed at a convenient location (such as a restaurant, a cafe, or a library). Multiple Cloudlets may form a distributed computing platform, which can further extend the available resources for mobile devices~\cite{li2001computation}. As the Cloudlet is just one hop away from the users' mobile devices, it improves the QoS with low communication delay and high bandwidth utilization.

In detail, Cloudlet has three main features as follows.
\subsubsection{Soft State}
Cloudlet can be regarded as a small cloud computing center located at the edge of the network. Therefore, as the server end of the application, the Cloudlet generally needs to maintain state information for interacting with the client. However, unlike Cloud, Cloudlet does not maintain long-term state information for interactions, but only temporarily caches some state information. This reduces much of the burden of Cloudlet as a lightweight cloud.

\subsubsection{Rich Resources}
Cloudlet has sufficient computing resources to enable multiple mobile users to offload computing tasks to it. Besides, Cloudlet also has stable power supply so it doesn't need to worry about energy exhaustion.

\subsubsection{Close to Users}
Cloudlets are deployed at those places where both network distance and physical distance are short to the end user, making it easy to control the network bandwidth, delay and jitter. Besides, the physical proximity ensures that the Cloudlet and the user are in the same context (e.g., the same location), based on which customized services (e.g., the location-based service) could be provided.

To further promote Cloudlet, CMU built up an open edge computing alliance, with Intel, Huawei and other companies~\cite{OpenEdgeCompute}, to develop standardized APIs for Cloudlet-based edge computing platforms. Currently, the alliance has transplanted OpenStack to the edge computing platform, which enables distributed Cloudlet control and management via the standard OpenStack APIs~\cite{ha2015openstack++}. With recent development of the edge computing, the Cloudlet paradigm has been widely adopted in various applications, e.g., cognitive assistance system~\cite{ha2013just,ha2014towards}, Internet of Things data analysis~\cite{satyanarayanan2015edge}, and hostile environments~\cite{satyanarayanan2013role}. 

Unlike the cloud, cloudlets are deployed on the edge of the network and serve only nearby users. Cloudlet supports application mobility, allowing devices to switch service requests to the nearest cloudlet during the mobile process. As shown in Fig.~\ref{fig:Cloudlet}, Cloudlet supports for application mobility relying on three key steps.

\textit{1) Cloudlet Discovery:} 
Mobile devices can quickly discover the available Cloudlets around them, and choose the most suitable one to offload tasks.

\textit{2) VM Provisioning:}
Configuring and deploying the service VM that contains the server code on the cloudlet so that it is ready to be used by the client.

\textit{3) VM Handoff:}
Migrating the VM running the application to another cloudlet.

\subsection{CloudPath}

\begin{figure}[!tb]
\begin{center}
\includegraphics[width=0.4\textwidth]{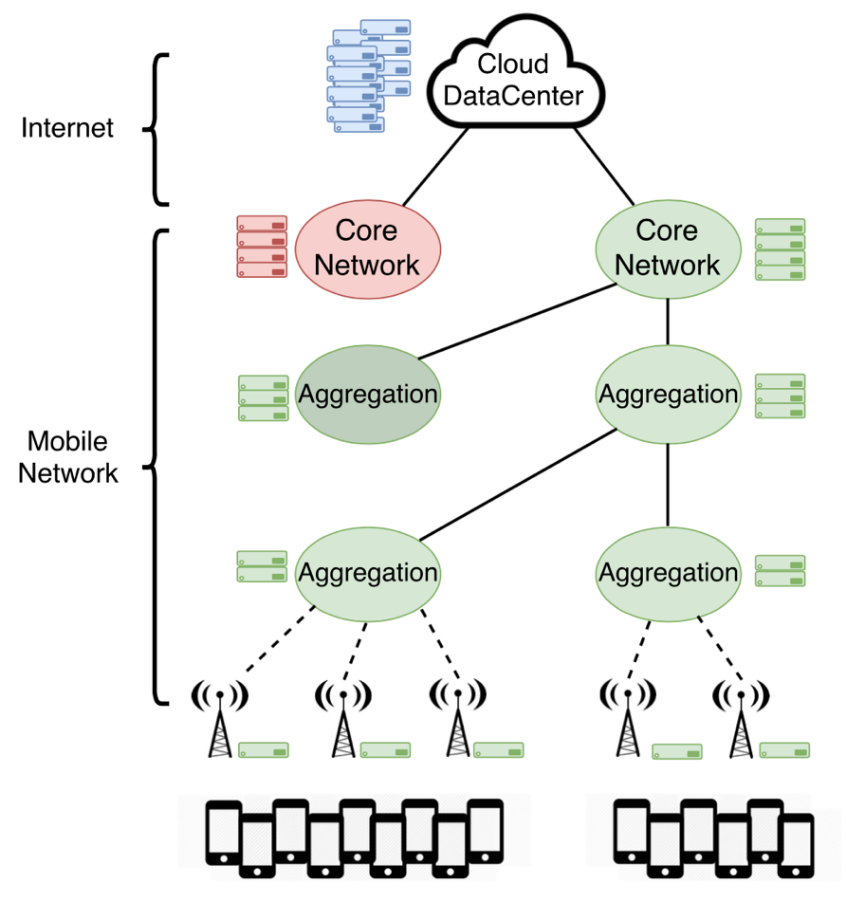}
\caption{CloudPath Architecture~\cite{cloudpath}.}\label{fig:CloudPath}
%\vspace{-0.20in}
\end{center}
\end{figure}

CloudPath~\cite{cloudpath} is an edge computing system proposed by the University of Toronto. In such a system, diverse resources like computing and storage are provided along the path from the user device to the cloud data center. It supports on-demand allocation and dynamic deployment of the multi-level architecture. The main idea of CloudPath is to implement the so-called ``path computing'', such that it can reduce the response time and improve the bandwidth utilization, compared with the conventional cloud computing. 

As illustrated in Fig.~\ref{fig:CloudPath}, the bottom layer of CloudPath is user devices, and the top layer is the cloud computing data center. The system reassigns those tasks of the data centers along the path (for path computing) to support different types of applications, such as IoT data aggregation, data caching services, and data processing services. Developers can select an optimal hierarchical deployment plan for their services by considering the factors such as cost, delay, resource availability and geographic coverage. Path computing builds a multi-tiered architecture, and from the top (traditional data center) to the bottom (user terminal equipment), the device capability becomes weaker, while the number of devices gets larger. On the premise of clear separation of computing and states, CloudPath extends the abstract shared storage layer to all data center nodes along the path, which reduces the complexity of third-party application development and deployment, and meanwhile keeps the RESTful development style.

The CloudPath application consists of a set of short-cycle and stateless functions that can be quickly instantiated at any level of the CloudPath framework. Developers either tag functions to specify where (such as edges, cores, clouds, etc.) their codes run, or tag performance requirements (such as response latency) to estimate the running location. %For examples, the identity authentication module can be run at any location; the face detection module needs to run on the device which provides response delay less than $10ms$. Furthermore, CloudPath provides a distributed eventual consistent storage service for reading and writing. The storage service automatically backs up the application's state to ensure minimum access latency and bandwidth overhead when using multiple levels of data centers. 
CloudPath does not migrate a running function/module, but supports service mobility by stopping the current instance and re-starting a new one at the expected location.
%Overall, CloudPath groups data centers at all levels along the path into a tree topology with underlying mobile networks and Internet coverage. The data center is called the CloudPath node, and a new node can be put to any layer of the tree. Note that the resources, capabilities, and quantity of different CloudPath nodes can be different. 
Each CloudPath node usually consists of the following six modules.
%\begin{itemize}
\textit{1) PathExecute:}   implements a serverless cloud container architecture that supports lightweight stateless application functions/modules.
\textit{2) PathStore:} provides a distributed eventual consistent storage system that transparently manages application data across nodes. \nop{PathStore is also used in PathDeploy and PathRoute to get the application program and routing information.}
\textit{3) PathRoute:} transmits the request to the most appropriate CloudPath node, according to the information such as user's location in the network, application preferences, or system status.
\textit{4) PathDeploy:} dynamically deploys and removes applications on CloudPath nodes based on application preferences and system policies.
\textit{5) PathMonitor:} provides real-time monitoring and historical data analysis function to applications and CloudPath nodes. It collects the metrics of other CloudPath modules on each node through the PathStore and presents the data to users using web pages.
\textit{6) PathInit:} is an initialization module in the top-level data center node and can be used to upload applications/services to the CloudPath.
%\end{itemize}

\subsection{PCloud}

\begin{figure}[!tb]
\begin{center}
\includegraphics[width=0.4\textwidth]{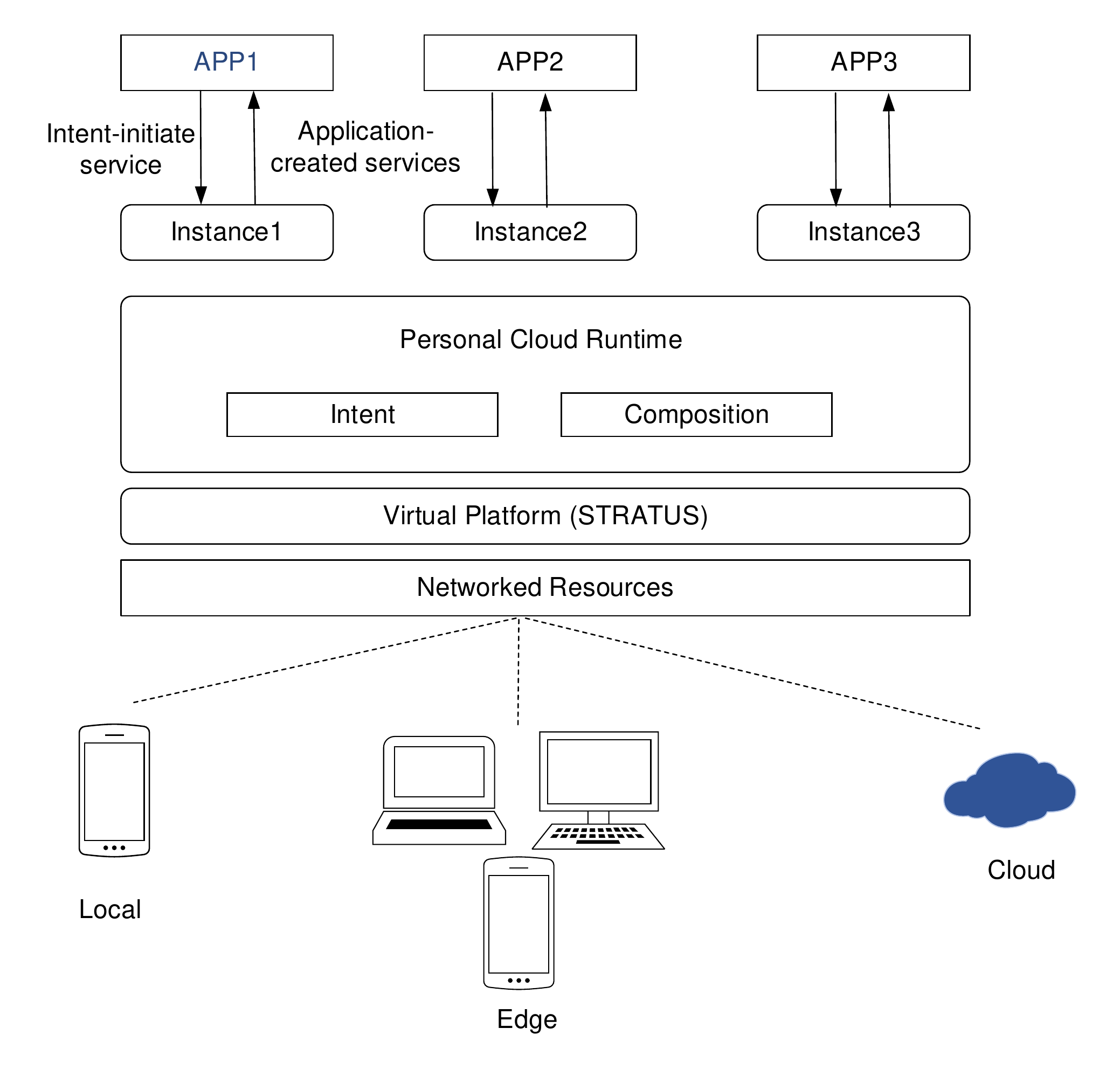}
\caption{PCloud Architecture~\cite{jang2014personal}.}\label{fig:PCloud}
%\vspace{-0.20in}
\end{center}
\end{figure}

PCloud~\cite{jang2014personal} integrates the edge computing and storage resources with those at the cloud to support seamless mobile services. The architecture of PCloud is shown in Fig.~\ref{fig:PCloud}. Specifically, these resources are virtualized through a special virtualization layer named STRATUS~\cite{jang2011stratus}, and form a distributed resource pool that can discover new resources and monitor resource changes. With the resource pool, the runtime mechanism is responsible for resource application and allocation. Through a resource description interface, the runtime mechanism selects and combines appropriate resources based on the requirements of specified applications. After the resources are combined, it generates a new instance to provide corresponding services for external applications, according to the resource access control policy. (Note that, the computing resources of the newly generated instance may come from multiple devices, which is equivalent to one integrated computing device for the external applications.) Thus, an application program is actually a combination of services running on the PCloud instance. For example, a media player application can be a combination of the storage, decoding and playing services. These services could be local or remote but are transparent to the applications. Furthermore, the PCloud system also provides basic system services, such as permission management and user data aggregation, to control the resources access of other users.

In the actual operation, the mobile application describes the required resources to the PCloud through interfaces. The PCloud will find out the optimal resource configuration by analyzing the description and the currently available resources, and then generates an instance to provide corresponding services for the application. %The resource evaluation metrics mainly include factors such as computing power and network delay. If the required resources include input/output devices, then the factors such as screen size and resolution may also be included.
PCloud integrates edge resources with cloud resources so that they can complement each other. The abundant resources from the cloud can make up for the lack of computing and storage capabilities at the edge; meanwhile, due to the physical proximity, edge devices can provide low-latency services to the user that cloud cannot offer. In addition, PCloud also enhances the availability of the entire system and can choose alternate resources when encountering network and equipment failures.

%Similar to PCloud's idea of combining edge resources with cloud resources, CoTWare~\cite{J2017cotware} focuses on supporting large-scale Internet of Things applications. It mainly provides two types of services: core services (such as geographic location services, security services, service calls, and other common system-level services) and environmental services (responsible for obtaining services provided by the cloud and edge devices, such as sensors, cameras, vehicles, etc.). As a user-centric edge computing platform, Femtoclouds~\cite{habak2017workload} allows tasks to be completed close to the user. Besides micro-servers, Femtoclouds can also be supported by small clusters formed by edge devices at the same locations (such as stations, classrooms, etc.). In addition, Femtoclouds need to ensure the communication with the remote cloud which is responsible for task management and resource tracking.

\subsection{ParaDrop}

\begin{figure}[!tb]
\begin{center}
\includegraphics[width=0.45\textwidth]{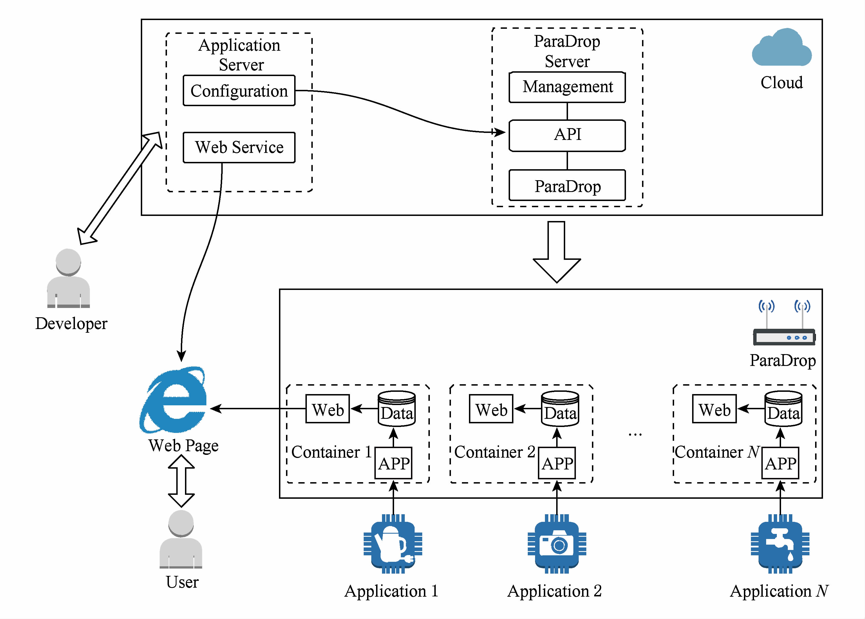}
\caption{ParaDrop System~\cite{paradrop}.}\label{fig:ParaDrop}
%\vspace{-0.20in}
\end{center}
\end{figure}

ParaDrop~\cite{paradrop} is developed by the WiNGS Laboratory at the University of Wisconsin-Madison. It is an edge computing framework that makes the computing/storage resources close to mobile devices and data sources available to the third party developers. Its goal is to bring intelligence to the network edge in a friendly way. 

ParaDrop upgrades the existing access point to an edge computing system, which supports applications and services like a normal server. To isolate applications under the multi-tenancy scenario, ParaDrop leverages the lightweight container virtualization technique. As Fig.~\ref{fig:ParaDrop} shows, the ParaDrop server (in the cloud) controls the deployment, starting and deletion of the applications. It provides a group of APIs, via which the developer can monitor and manage the system resources and configure the running environment. The web UI is also provided, through which the user can directly interact with the applications.

The design goals of ParaDrop include three aspects: multi-tenancy, efficient resource utilization and dynamic application management. To achieve these goals, the container technology is applied to manage the multi-tenancy resources separately.\nop{Besides, the flexible APIs helps developers allocate specific resources to each application.} As the resources of the edge devices are very limited, compared with the virtual machine, the container consumes less resources and would be more suitable for delay sensitive and high I/O applications. Moreover, as applications running in the container, ParaDrop can easily control their startup and revocation.

ParaDrop is mainly used for IoT applications, especially IoT data analysis. Its advantages over traditional cloud system can be summarized as follows: a) since sensitive data can be processed locally, it protects the users’ privacy; b) WiFi access point is only one hop away from the data source, leading to low network delay and stable connection; c) only user requested data are transmitted to the equipment through the Internet, thus cutting down the total traffic amount and saving the bandwidth of the backbone network; d) the gateway can obtain the location information of the edge devices through radio signals (e.g., the distance between devices, and the location of the specific device), which facilitates the location-aware services; e) when edge devices cannot be connected to the Internet, the edge service can still work. \nop{In a word, ParaDrop is well developed, the software system is fully open source, and the hardware devices that support ParaDrop system are also on the market.}

\subsection{SpanEdge}

\begin{figure}[!tb]
\begin{center}
\includegraphics[width=0.5\textwidth]{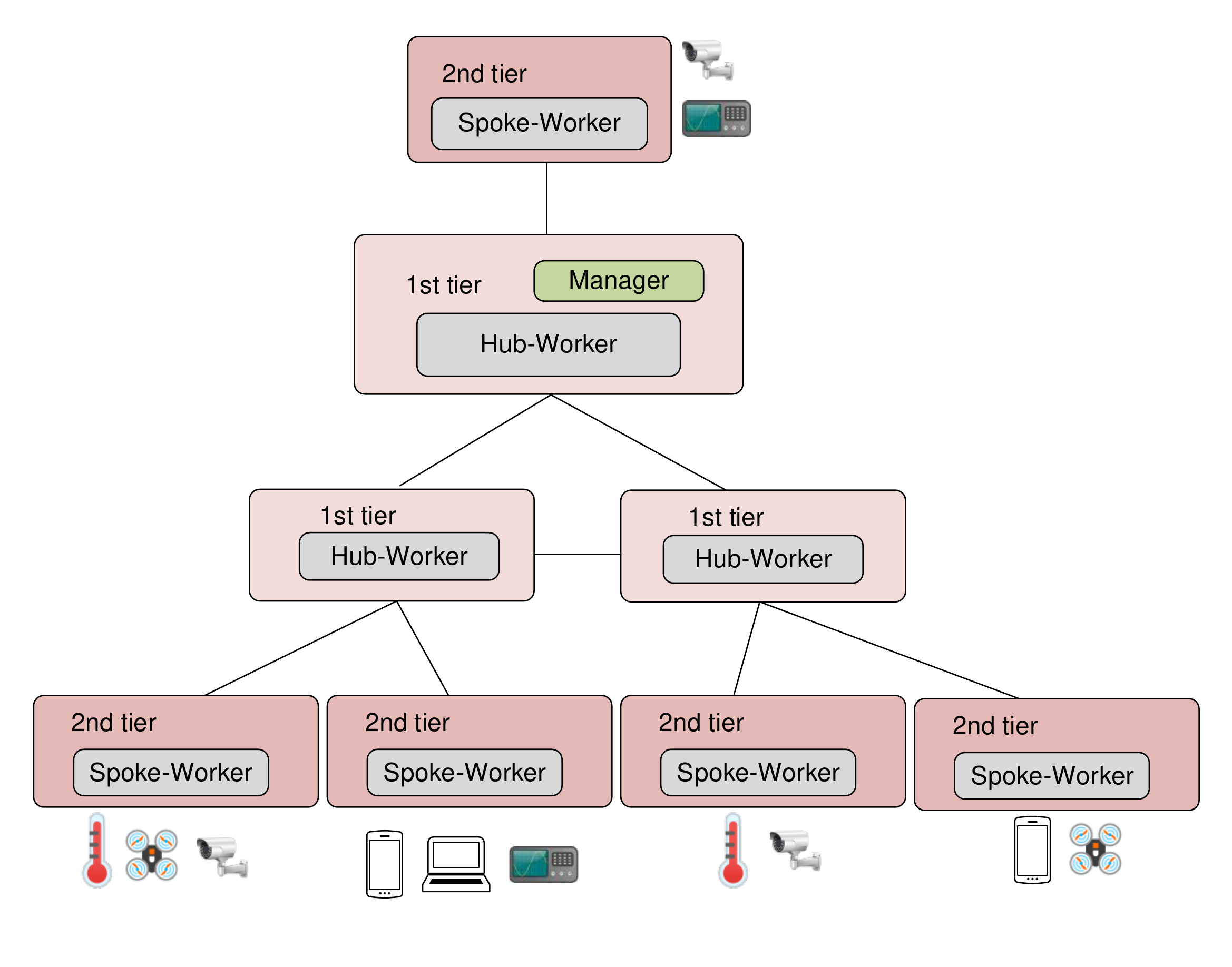}
\caption{SpanEdge Architecture~\cite{spanedge}.}\label{fig:SpanEdge}
%\vspace{-0.20in}
\end{center}
\end{figure}

Streaming processing is one important type of applications in edge computing, where data is generated by various data sources in different geographical locations and continuously transmitted in the form of streams. Traditionally, all raw data is transmitted over the WAN to the data center server, and stream processing systems, such as Apache Spark and Flink, are also designed and optimized for one centralized data center. However, this approach cannot effectively handle the huge data generated by a lot of devices at the edge of the network, and the situation is even worse when the applications require low latency and predictability. SpanEdge~\cite{spanedge} is a research project of the Royal Institute of Technology in Sweden. It unifies the cloud central node and the near-edge central node, reduces network latency in WAN connections, and provides a programming environment that allows the program to run near the data source. Developers can focus on developing streaming applications without considering where the data sources are located and distributed.

The data center in SpanEdge is composed of two levels: the cloud data center is the first level and the edge data center (such as the operator's cloud, Cloudlet, or Fog) is the second level. Partial streaming processing tasks run on the edge central nodes to reduce latency and boost performance. SpanEdge uses the master-worker architecture (as shown in Fig.~\ref{fig:SpanEdge}) with one manager and multiple workers. The manager collects the streaming processing requests and assigns tasks to the workers. Workers mainly consist of cluster nodes whose primary responsibility is to execute tasks. There are two types of workers: hub-worker (first level) and spoke-worker (second level). \nop{They have no functional differences.} The network transmission overhead and latency are related to the geographical location and network connection status between workers. The communication in SpanEdge is also divided into a system management communication (worker-manager) and a data transmission communication (worker-worker). System management communication aims to schedule tasks between managers and workers, and data transmission communication takes care of the data flow in each task. Each worker has an agent that handles system management operations, such as sending and receiving management information, monitoring compute nodes to ensure that they are running normally, and periodically sending heartbeat messages to the manager to ensure immediate recovery when the task fails.

SpanEdge allows developers to divide the tasks into local ones and global ones. Local tasks should run on the node near the data source and provide only part of the required data; global tasks are responsible for further processing the results of local tasks and aggregating all results. SpanEdge creates a copy of the local task on each spoke-worker which has all corresponding data sources. If the data is insufficient, the task is dispatched to the hub-worker. The global task runs on a separate hub-worker, and the scheduler selects the optimal hub-worker based on the network delay (i.e. the distance from the spoke-worker in the network topology).

\subsection{Cloud-Sea Computing Systems}

\begin{figure}[!tp]
	\centering
	\includegraphics[width=0.9\linewidth]{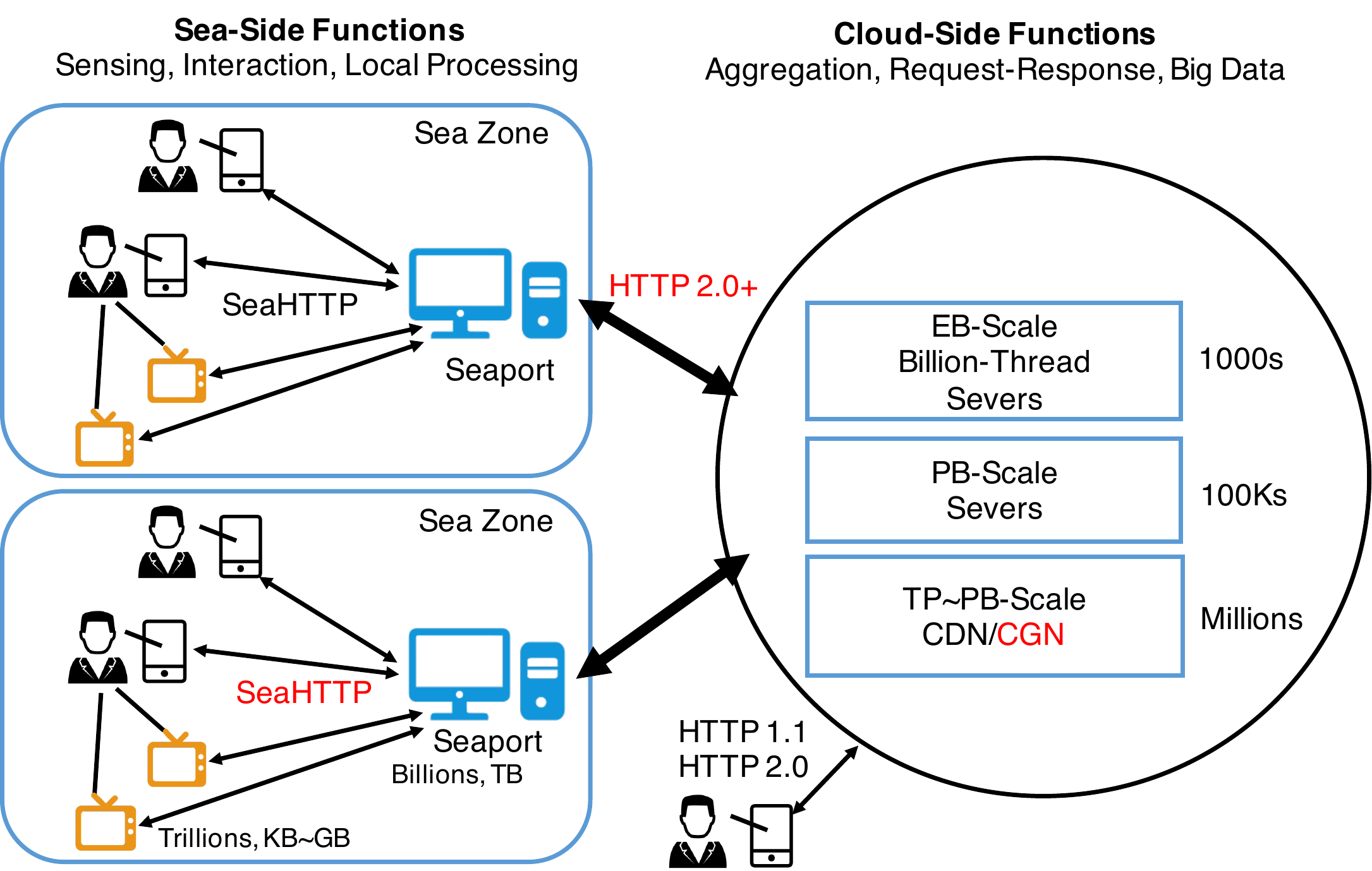}
	\caption{Cloud-Sea Computing Model.}
	\label{fig:Cloud-Sea}
\end{figure}

Cloud-Sea Computing Systems project~\cite{xu2014cloud} is a main research thrust of the Next Generation Information and Communication Technology initiative (the NICT initiative) and a 10-year strategic priority research initiative, launched by the Chinese Academy of Science in 2012. The NICT initiative aims to address the three major technology challenges in the coming Zettabyte era: i) improving the performance per watt by $1000$ times, ii) supporting more applications from the human-cyber-physical ternary computing, and iii) enabling transformative innovations in devices, systems and applications, while without polluting beneficial IT ecosystems.

In the cloud-sea computing system, ``cloud'' refers to the datacenters and ``sea'' refers to the terminal side (the client devices, e.g., human-facing and physical world facing subsystems). The design of the project can be depicted from three levels: the overall systems architecture level, the datacenter server and storage system level, and the processor chip level. The project contains four research components: a computing model called REST 2.0 which extends the representational state transfer (REST)~\cite{fielding2000architectural} architectural style of Web computing to cloud-sea computing, a three-tier storage system architecture capable of managing ZBs of data, a billion-thread datacenter server with high energy efficiency, and an elastic processor aiming at energy efficiency of one trillion operations per second per watt.

As shown in Fig.~\ref{fig:Cloud-Sea}, the cloud-sea computing model includes sea-side functions and cloud-side functions. The sea zone is expanded from the traditional cloud client, e.g., a home, an office, a factory manufacturing pipeline. There can be multiple client devices inside a sea zone, and each device can be human facing or physical world facing. In a sea zone, there is a special device (like a home datacenter or a smart TV set) designated as the seaport for three purposes: i) a gateway interfacing the sea zone to the cloud, ii) a gathering point of information and functionalities inside a sea zone, and iii) a shield protecting security and privacy of the sea zone. A device inside a sea zone does not communicate to the cloud directly, but through the seaport, either implicitly or explicitly. The SeaHTTP, a variant based on HTTP $2.0$, is the widely-used protocol in sea zones of the cloud-sea system to connect with the cloud.

The cloud-sea computing model has four distinct features. 

\textit{1) Ternary computing via sea devices:} 
Human and physical world entities interface and collaborate with the cyberspace through the sea side. For example, users can leverage a smart phone application to read and control a sensor device in a home through an Internet application service.

\textit{2) Cooperation with locality:} 
A specific network computing system will partition its functions between the sea side and the cloud side. Sea-side functions include sensing, interaction, and local processing while cloud-side functions include aggregations, request-response, and big data processing.

\textit{3) Scalability to ZB and trillion devices:} 
This future Net will collectively need to support trillions of sea devices and to handle ZBs of data.

\textit{4) Minimal extension to existing ecosystems:} 
The REST $2.0$ cloud-sea computing architecture attempts to utilize existing Web computing ecosystems as much as possible. 

Overall, the cloud-sea computing model is proposed to migrate the cloud computing function to the sea side, and it focuses more on the devices at the ``sea'' side and the data at the ``cloud'' side. Typical edge computing is more generalized and may care about any intermediate computing resources and network resources between the ``sea'' and the ``cloud''. The researches in cloud-sea computing (e.g., energy efficient computing and elastic processor designing) are consultative for the edge computing.

\subsection{Cachier and Precog}
\label{subsec:Cachier_and_Precog}

\begin{figure}[!tb]
\begin{center}
\includegraphics[width=0.5\textwidth]{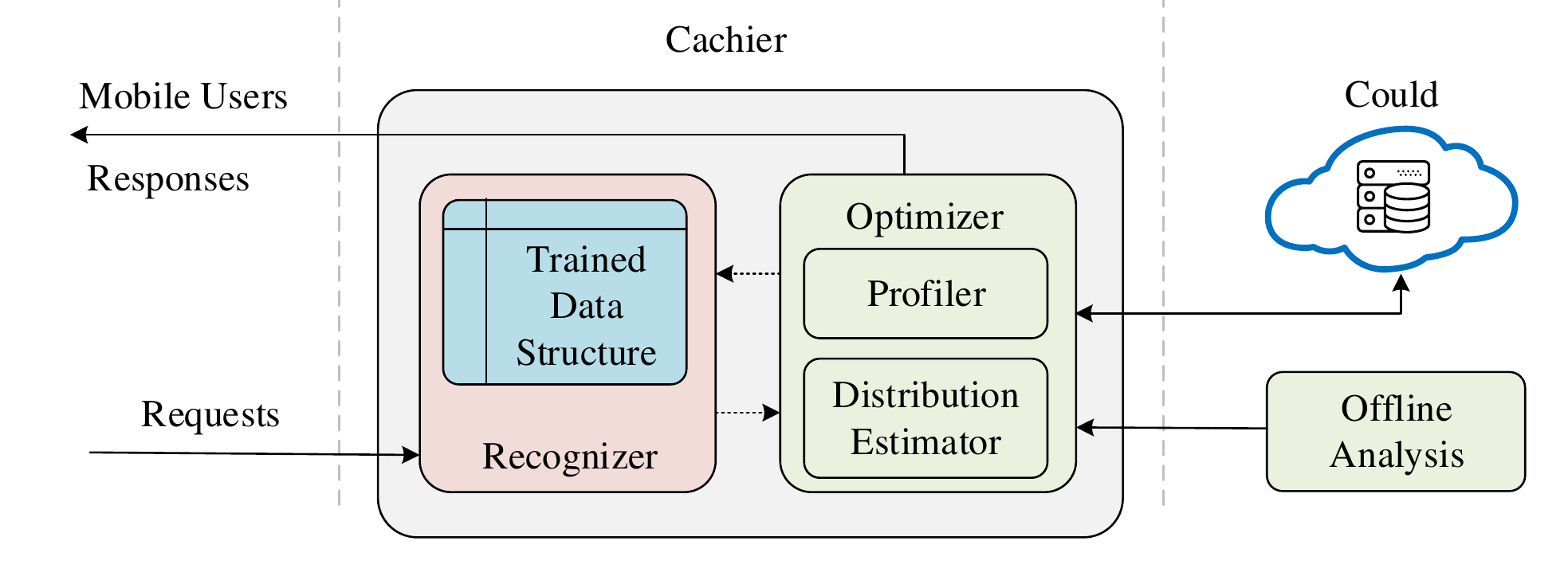}
\caption{Cachier System~\cite{cachier}.}\label{fig:Cachier}
%\vspace{-0.20in}
\end{center}
\end{figure}

Cachier~\cite{cachier} and Precog~\cite{precog} are two edge caching systems that proposed by the researchers from Carnegie Mellon University for image recognition. %With the development of wearable devices and artificial intelligence technology, more and more recognition and awareness mobile applications appear, such as face recognition, speech recognition, landmark recognition, augmented reality, and so on. Such applications usually have soft real-time features, that is, if the recognition time is too long, the user's focus may have shifted to the next object, and the previous identification request and results are meaningless. Therefore, response time is the most critical performance indicator which should be controlled within a few hundred milliseconds. However, such applications are usually CPU-intensive, and the accurate results are obtained after searching a large number of candidates with the trained recognition model. 
Recognition applications have strict response time requirements, while the computation is huge due to the model complexity and dataset size. With edge computing, we can leverage the computing resource of the edge nodes to process the matching, so the network delay can be reduced.
%It is not feasible to perform such huge computation on end devices (e.g., a smartphone). The traditional approach is to collect data (e.g., video, image and audio) on the end device, and then send them to the cloud for feature extraction and matching. In the edge computing environment, one solution is to use the computing resource of the edge nodes to process the matching so that the network delay between the client and the cloud can be reduced. 
Considering that edge nodes mainly provide services to nearby users, the spatio-temporal characteristics of service requests can be leveraged. From the perspective of caching, Cachier system proposes that edge nodes can be used as ``computable cache devices'', which can cache recognition results, reduce matching dataset size and improve response time. By analyzing the response delay model and requests' characteristics, the system can dynamically adjust the size of the dataset on the edge nodes according to the environment factors, thus ensuring optimal response time. %Cachier system illustrates the overall system architecture and operation mechanism by taking the image recognition application as an example, and it can also be extended to other recognition applications.

As Fig.~\ref{fig:Cachier} shows, Cachier consists of the Recognizer Module, the Optimizer Module, and the Offline Analysis Module. %Taking the Least Frequently Used (LFU) cache replacement strategy as an example, the Cachier system can modify the LFU to a dynamic adaptive strategy to reduce the figure recognition response time. 
The Recognizer Module is responsible for analyzing and matching the received figures according to the cached training data and model. If there is a match, the result will be directly returned to the user, otherwise the figure will be transmitted to the cloud for recognition. %Using the LFU cache strategy, the cached data is the top $k$ objects of recent recognitions. Therefore, 
The distribution estimator in the Optimizer Module can use the maximum a posteriori estimation to predict the request distribution. Secondly, given the classification algorithm and training dataset, the cache searching delay and the cache accuracy can also be calculated by the Offline Analysis Module. At last, the Profiler sub-module is responsible for estimating the network delay and cloud latency incurred by cache misses. It measures and records the delay under corresponding distance in real time, and uses the moving average filter to remove the noise data. By taking such information into the delay expectation time model, Cachier is able to calculate the optimal cache size, and then adjusts the cache on the edge nodes accordingly. \nop{In addition, the Cachier system also provides other optimizations, such as hot start and feature request waiting queues.}

\begin{figure}[!tb]
\begin{center}
\includegraphics[width=0.5\textwidth]{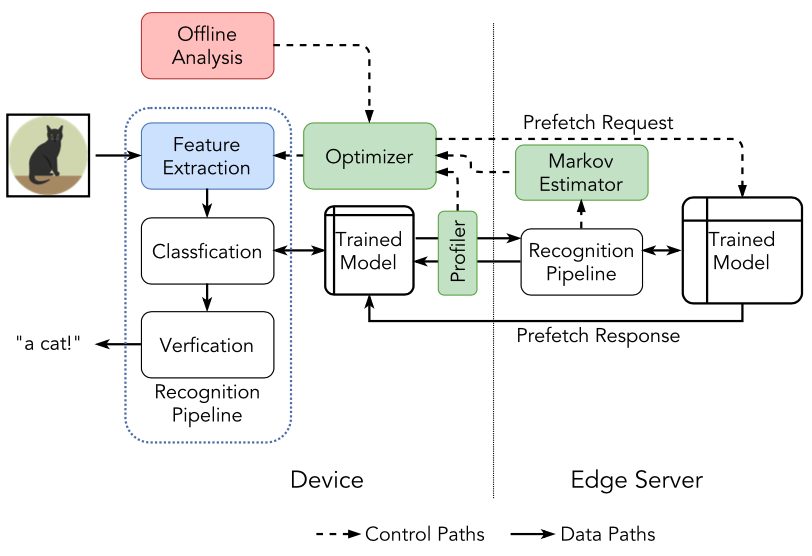}
\caption{Precog System~\cite{precog}.}\label{fig:Precon}
%\vspace{-0.20in}
\end{center}
\end{figure}

Precog is an extension of Cachier. The cached data is not only on the edge nodes, but also on the end devices which are used for selection calculation in order to reduce the image migration between the edge nodes and the cloud. Based on the prediction model, Precog prefetches some of the trained classifiers, uses them to recognize the images, and cooperates with the edge nodes to efficiently complete the tasks. %With Cachier, the total response time of the user requests can be reduced by shortening the delay from the edge node to the cloud, while it cannot solve the last mile problem, i.e., the delay from the end devices to the edge. In contrast, 
Precog pushes computing capability to the end device and leverages the locality and selectivity of the user requests to reduce the last mile delay. For the end device, if the cache system uses the same cache replacement policy as the edge node applies, it will lead to a large number of forced misses. For a single user, the device only has the request information of the user, and usually the user will not identify the same object multiple times in the near future. Therefore, Precog constructs a Markov model using the request information from the edge node to describe the relationship among the identified objects, and then predicts the potential future requests. %Besides, compared with Cachier, 
Precog improves the delay expectation time model by considering the network state, device capabilities and the prediction information from edge nodes. %This model is used to select the appropriate data as caches on the end device. \nop{In addition, Precon also proposes to dynamically modify the feature extraction module based on the running environment to further optimize response time.}
The system architecture of Precog is illustrated in Fig.~\ref{fig:Precon}. We can see that it is mainly composed of the feature extraction module, the offline analysis module, the optimizer module and the profiler module. The edge node is mainly responsible for the construction of Markov model. Based on the offline analysis results, the Markov model prediction results, and the network information provided by the profiler module, the optimizer module determines the number of feature to be extracted as well as the data to be cached on the end device.

\subsection{FocusStack}

\begin{figure}[!tb]
\begin{center}
\includegraphics[width=0.5\textwidth]{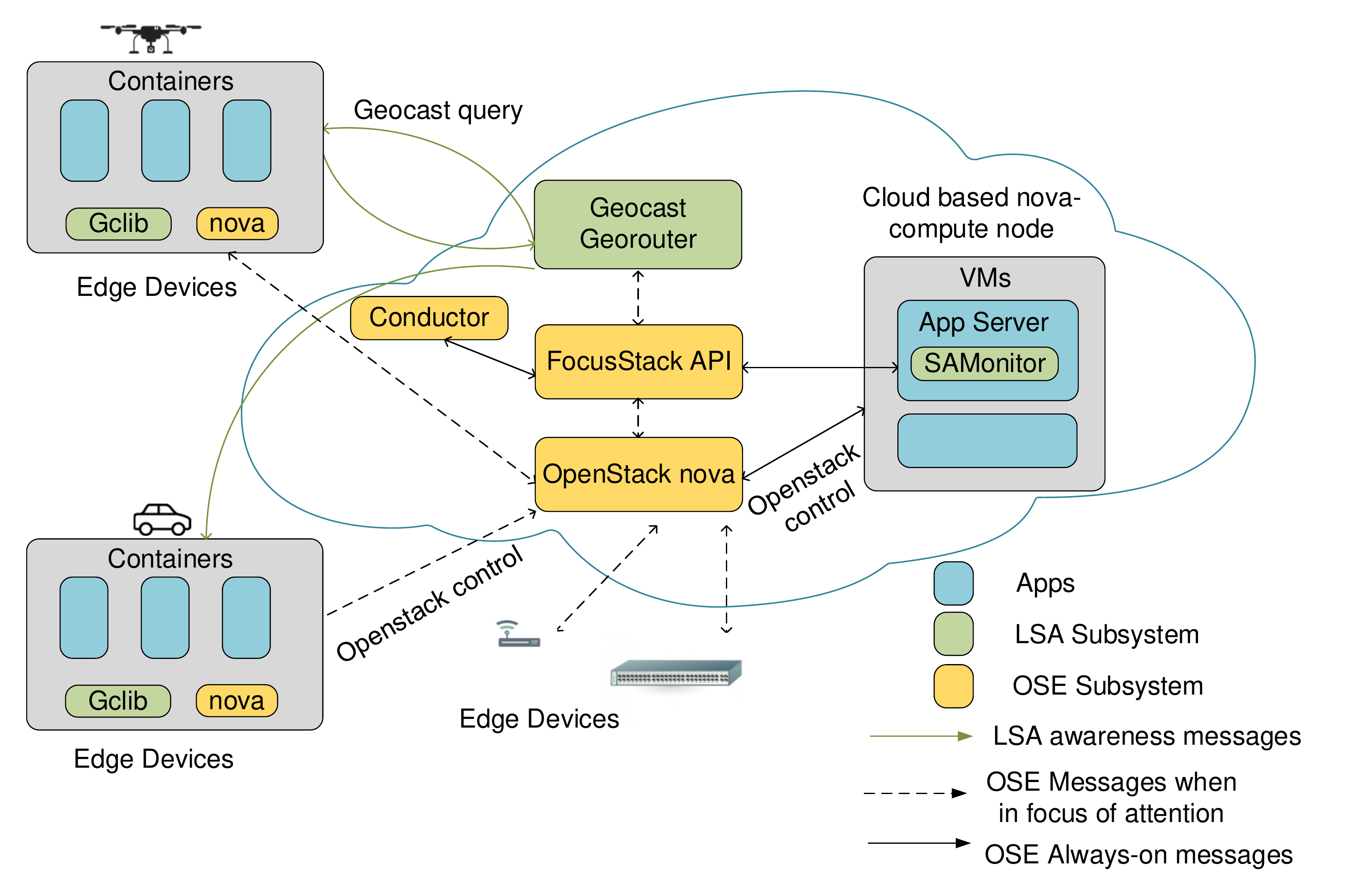}
\caption{FocusStack System~\cite{focusstack}.}\label{fig:FocusStack}
%\vspace{-0.20in}
\end{center}
\end{figure}

FocusStack~\cite{focusstack} is developed by AT\&T Labs, which supports the deployment of complex applications on a variety of potential IoT edge devices. Although the resources, such as computing, power consumption, connectivity, are limited on edge devices, they have the nature of mobility. Hence, it is very important to have a system that can discover and organize available edge resources. FocusStack is such a system that can discover a set of edge devices with sufficient resources, deploy applications and run them accordingly. Thus, the developers can focus more on the design of applications than on how to find and track edge resources.

FocusStack consists of two parts (as shown in Fig.~\ref{fig:FocusStack}): i) Geocast system, which provides location-based situational awareness (LSA) information, and ii) OpenStack extension (OSE), which is responsible for deploying, executing, and managing the containers on the edge devices. FocusStack builds a hybrid cloud of edge devices (containers) and data center servers (virtual machines). When a user initiates a cloud operation through the FocusStack API (such as instantiating a container), the LSA subsystem analyzes the scope of the request based on the Geocast route and sends a resource list of geographic locations to the target area, and waits for the response from (online) edge devices that can satisfy the requirements. Then, the selected edge device runs the corresponding OpenStack operations with the help of the conductor module.

The situational aware subsystem enables one group of devices to monitor the survival status of each other, and each device can update the sensing information of other devices. %AT\&T Lab’s Geocast System (ALGS) is used to implement the subsystem. The address of a packet indicates a physical space, which means that the packet should be sent to all devices in that space. The packet address gives a circular area defined by latitude, longitude and radius. 
The geo-processing projection service is primarily intended to pass requests and reply messages between areas of interest, send out device control information (like drones) and broadcast location-based information. The service is based on a two-layer network, and data can be transmitted over a dedicated WiFi network or the Internet. The sender transmits the data packet to the geo-router server which tracks the location and metadata of each edge device, and then the data packet is sent to each device (including edge devices and a cloud device running a SAMonitor instance) in the specified area based on the address. Location and connection statuses are maintained by the geo-routing database (GRDB). The GCLib is a software framework that provides data acquisition services, and it runs on edge devices and cloud applications that use SAMonitor. %GCLib packs the data payload information and geographic information into a geocast packet and leverages the related protocol to send out the data packet. 
Any application or service in need of the state of the device in the area needs a SAMonitor component to communicate with the LSA subsystem. The application server makes a request to SAMonitor through the FocusStack API, and then SAMonitor builds a real-time graph of the current area and returns a list of available edge devices. The regional graph is sent to the conductor in the OSE, which is responsible for checking whether the devices are capable of running the tasks, whether the pre-defined policy rules are met, and so on. After that, the available edge device list is submitted to the application server and the server selects devices to be used. OSE manages and deploys the program through the OpenStack Nova API. The edge devices run a custom version of Nova Compute to interact with the local Docker to manage the container. The container on the edge devices supports all OpenStack services, including access to virtual networks and application-based granularity configuration.

\subsection{AirBox}

\begin{figure}[!tb]
\begin{center}
\includegraphics[width=0.5\textwidth]{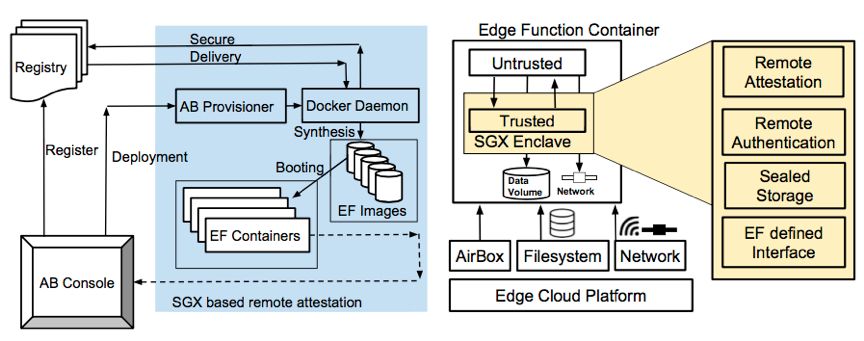}
\caption{AirBox Architecture~\cite{airbox}.}\label{fig:AirBox}
%\vspace{-0.20in}
\end{center}
\end{figure}

AirBox~\cite{airbox} is a secure, lightweight and flexible edge function system developed by Georgia Institute of Technology. It supports fast and flexible edge function loading and provides service security and user privacy protection. The Edge Function (EF) in AirBox is defined as a service that can be loaded on an edge node, and the software stack that supports the edge function is named Edge Function Platform (EFP). 

As shown in Fig.~\ref{fig:AirBox}, the AirBox consists of two parts: the AB console and the AB provisioner. The back-end service manager deploys and manages the EF on the edge nodes through the AB console. The AB provisioner which runs at the edge is responsible for providing dynamic EF seamlessly. The edge functions are implemented through system-level containers with minimal constraints on developers. %, but this method requires additional mechanisms to ensure the integrity of edge functions. They cannot rely on the trusted components on the edge system to ensure security, as the security of the edge device and the system itself are not guaranteed. Instead, 
Security is enhanced by using the hardware security mechanisms like Intel SGX. AirBox provides centrally controlled backend services in discovering edge nodes and registering them. The AB console is a web-based management system, which activates the docker start-up process on the edge node with AB provisioner. %In detail, by leveraging the container registration service, the docker daemon takes over the EF containers from the cloud server.

To ensure the security of the AirBox, the EF consists of a trusted part and an untrusted part. The untrusted part is responsible for all network and storage interactions. Based on OpenSGX APIs, AirBox provides four extensible APIs to implement secure communication and storage: Remote Attestation, Remote Authentication, Sealed Storage, and EF Defined Interface. AirBox supports a variety of edge features such as aggregation, buffering, and caching. %The Aggregation function collects multiple requests from users and sends them to the backend after filtering redundant ones. Particularly, the IoT hub on the edge node can aggregate the information sent by multiple sensors and periodically transmit effective data to the backend service, so the required bandwidth can be reduced. A buffering mechanism is applied to store the returned information from the backend server on the edge node and send it to the user in a suitable context environment, e.g., sending the push notifications of the pre-configured applications based on the user location information. Furthermore, caching mechanism is also utilized to temporarily store the returned data of one user and reuse it for the same request from other users.

\subsection{Firework}

Wayne State University's MIST Lab proposes a programming model for edge computing\textemdash Firework~\cite{zhang2016firework}. In the model of Firework, all services/functions are represented in a data view of datasets and functions which can be the results of processing with their own data or secondary processing with other services/datasets. A system consisting of nodes that apply the Firework model is called the Firework system. In the Firework system, instead of dividing nodes into edge nodes and cloud ones, they are all considered as Firework nodes. Through the mutual invocation of the Firework nodes, the data can be distributed and processed on each node. Along the path of data transmission, the edge nodes perform a series of calculations upon the data, thus forming a ``computational flow''. The beginning of the computational flow could be user terminals, edge servers close to the user, edge servers close to the cloud, or cloud nodes. 

In the Firework system, the data processing service is split into multiple sub-services, and the scheduling of the data processing is performed at the following two layers.

\textit{1) Same sub-service layer scheduling:} 
A Firework node can cooperate with another for the same sub-services in the surrounding area to achieve optimal response time. For idle Firework nodes, the system can schedule sub-service programs onto them, such that a cluster providing specific sub-service is formed dynamically and complete the service faster.

\textit{2) Computational flow layer scheduling:} 
Firework nodes along the computational flow can cooperate with each other and dynamically schedule execution nodes to achieve an optimal solution. For example, depending on the states of the network, the system can choose Firework nodes for service providing based on the nodes' locations (e.g., selecting those closest to the users).

\begin{figure}[!tb]
\begin{center}
\includegraphics[width=0.35\textwidth]{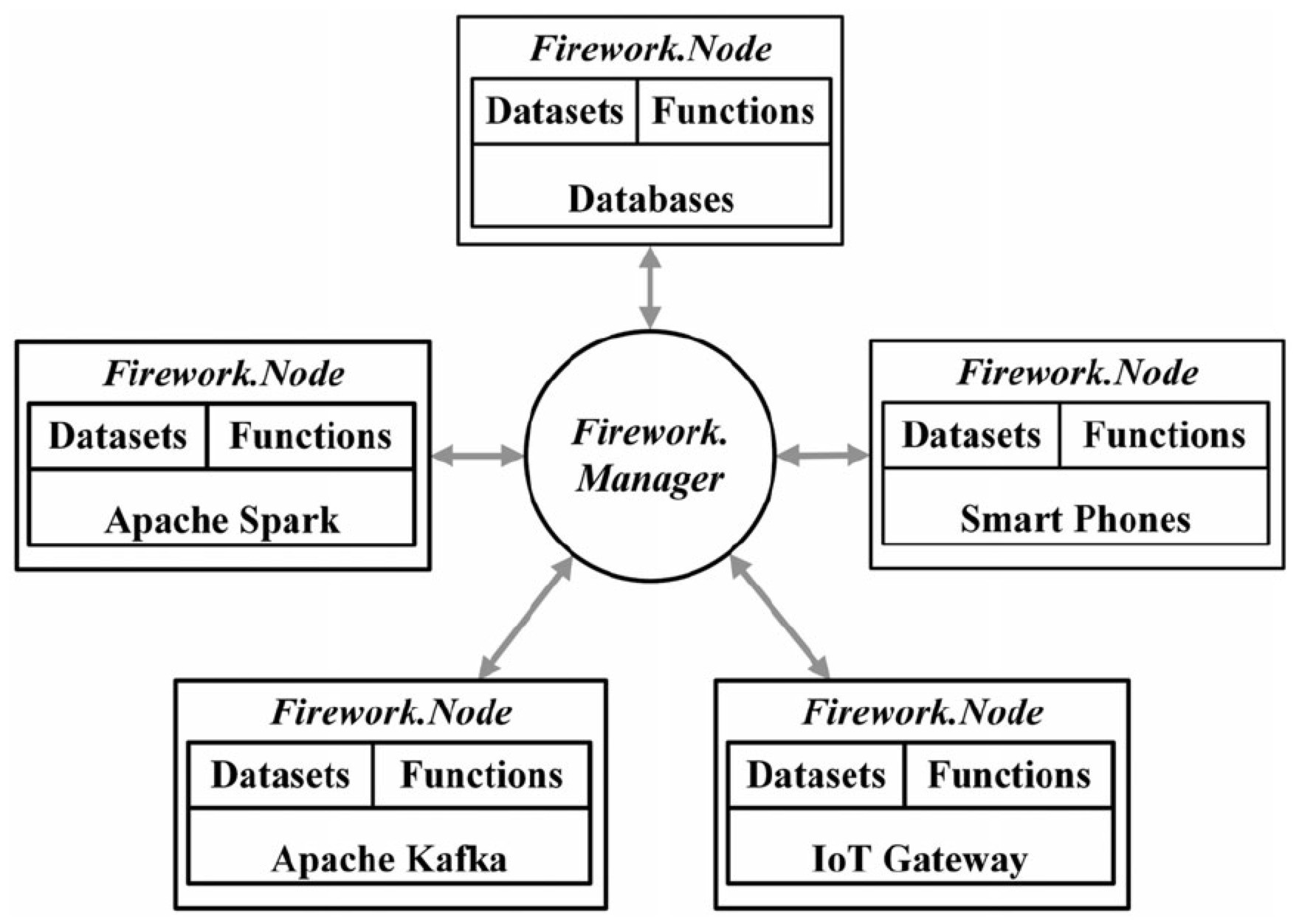}
\caption{An example of a Firework instance that consists of heterogeneous computing platforms~\cite{zhang2016firework}.}\label{fig:Firework}
%\vspace{-0.20in}
\end{center}
\end{figure}

As shown in Fig.~\ref{fig:Firework}, Firework divides the nodes into computing nodes and managers, depending on the type of service those nodes provide. In general, each node with the Firework model has three modules: the job management module, the actuator management module, and the service management module.

\begin{table*}[htb]
\centering
\caption{Summary of edge computing systems} %working in different application scenarios}
\label{tab:otherEdgeSystems}
\begin{tabular}{cccccc}
\hline
\textbf{\begin{tabular}[c]{@{}c@{}}Application\\ Scenarios\end{tabular}}              & \textbf{\begin{tabular}[c]{@{}c@{}}Edge\\ Computing\\ Systems\end{tabular}} & \textbf{\begin{tabular}[c]{@{}c@{}}End\\ Devices\end{tabular}}     & \textbf{Edge Nodes}                                                      & \textbf{\begin{tabular}[c]{@{}c@{}}Computation\\ Architecture\end{tabular}} & \textbf{Features/Targets} 
                                               \\ \hline
\multirow{10}{*}{\begin{tabular}[c]{@{}c@{}} General \\ Usage \\ Scenario \end{tabular}} & Cloudlet                                                                    & Mobile devices                                                               & Cloudlet                                                                    & \begin{tabular}[c]{@{}c@{}}Hybrid\\ (3-tier)\end{tabular}                   & \begin{tabular}[c]{@{}c@{}}Lightweight\\ VM migration\end{tabular}                              
\\ \cline{2-6}
                                                                                      & PCloud                                                               & Mobile devices  & \begin{tabular}[c]{@{}c@{}}Mobile devices,\\ local server, PC\end{tabular}                                             & \begin{tabular}[c]{@{}c@{}}Hybrid\\ (3-tier)\end{tabular}                   & \begin{tabular}[c]{@{}c@{}}Resource integration, \\ dynamic allocation\end{tabular} 
\\ \cline{2-6}
                             & ParaDrop                                                               & IoT devices  & Home gateway                                             & \begin{tabular}[c]{@{}c@{}}Hybrid\\ (3-tier)\end{tabular}                   & \begin{tabular}[c]{@{}c@{}}Hardware, \\ developer support\end{tabular}
\\ \cline{2-6}
                             & Cachier \& Precog                                                               & Mobile devices  & \begin{tabular}[c]{@{}c@{}}Mobile devices,\\ local server, PC\end{tabular}                                & \begin{tabular}[c]{@{}c@{}}Hybrid\\ (3-tier)\end{tabular}                   & \begin{tabular}[c]{@{}c@{}}Figure recognition, \\ identification \end{tabular}
\\ \cline{2-6}
                             & FocusStack                                                              & IoT devices  & Router, server                                 & \begin{tabular}[c]{@{}c@{}}Hybrid\\ (3-tier)\end{tabular}                   & \begin{tabular}[c]{@{}c@{}}Location-based info, \\ OpenStack extension \end{tabular}
\\ \cline{2-6}
                             & SpanEdge                                                              & IoT devices  & \begin{tabular}[c]{@{}c@{}}Local cluster,\\ Cloudlet, Fog \end{tabular}                                      & \begin{tabular}[c]{@{}c@{}}Hybrid\\ (2-tier)\end{tabular}                   & \begin{tabular}[c]{@{}c@{}}Streaming processing, \\ local/global task \end{tabular}
\\ \cline{2-6}
                             & AirBox                                                              & IoT devices  & \begin{tabular}[c]{@{}c@{}}Mobile devices,\\ local server, PC\end{tabular}                                       & \begin{tabular}[c]{@{}c@{}}Hybrid\\ (3-tier)\end{tabular}                   & Security
\\ \cline{2-6}
                             & CloudPath                                                              & Mobile devices  & 
                             \begin{tabular}[c]{@{}c@{}}Multi-level \\data centers\end{tabular}& \begin{tabular}[c]{@{}c@{}}Hybrid\\ (multi-tier)\end{tabular}                   & Path computing
\\ \cline{2-6}
                             & Firework                                                              & Firework.Node  & 
                             Firework.Node & \begin{tabular}[c]{@{}c@{}}Two-layer\\ scheduling\end{tabular}                   & Programming model
\\ \cline{2-6}
                             & Cloud-Sea                                                              & Sea  & 
                             Seaport & \begin{tabular}[c]{@{}c@{}}Hybrid\\ (3-tier)\end{tabular}                      & \begin{tabular}[c]{@{}c@{}}Minimal extension,\\transparency\end{tabular}
\\ \hline
\multirow{2}{*}{\begin{tabular}[c]{@{}c@{}}Vehicular\\ Data\\ Analytics\end{tabular}} & OpenVDAP                                                                    & CAVs                                                               & XEdge                                                                    & \begin{tabular}[c]{@{}c@{}}Hybrid\\ (2-tier)\end{tabular}                   & \begin{tabular}[c]{@{}c@{}}General\\ platform\end{tabular}                               \\ \cline{2-6}
                                                                                      & SafeShareRide                                                               & \begin{tabular}[c]{@{}c@{}}Smartphones\\ and vehicles\end{tabular} & Smartphones                                                              & \begin{tabular}[c]{@{}c@{}}Hybrid\\ (2-tier)\end{tabular}                   & \begin{tabular}[c]{@{}c@{}}In-vehicle\\ security\end{tabular}                            \\ \hline
\multirow{2}{*}{\begin{tabular}[c]{@{}c@{}}Smart\\ Home\end{tabular}}                 & Vigilia                                                                     & \begin{tabular}[c]{@{}c@{}}Smart\\ home devices\end{tabular}       & Hubs                                                                     & \begin{tabular}[c]{@{}c@{}}Edge\\ only\end{tabular}                         & \begin{tabular}[c]{@{}c@{}}Smart\\ Home Security\end{tabular}                            \\ \cline{2-6}
                                                                                      & HomePad                                                                     & \begin{tabular}[c]{@{}c@{}}Smart\\ home devices\end{tabular}       & Routers                                                                  & \begin{tabular}[c]{@{}c@{}}Edge\\ only\end{tabular}                         & \begin{tabular}[c]{@{}c@{}}Smart\\ Home Security\end{tabular}                            \\ \hline
\multirow{3}{*}{\begin{tabular}[c]{@{}c@{}}Video\\ Stream\\ Analytics\end{tabular}}   & LAVEA                                                                       & $\sim$                                                                  & $\sim$                                                                        & \begin{tabular}[c]{@{}c@{}}Edge\\ or cloud\end{tabular}                     & \begin{tabular}[c]{@{}c@{}}Low\\ latency response\end{tabular}                           \\ \cline{2-6}
                                                                                      & VideoEdge                                                                   & Cameras                                                            & \begin{tabular}[c]{@{}c@{}}Cameras and\\ private\\ clusters\end{tabular} & \begin{tabular}[c]{@{}c@{}}Hybrid\\ (3-tier)\end{tabular}                   & \begin{tabular}[c]{@{}c@{}}Resource-accuracy\\ tradeoff\end{tabular}                     \\ \cline{2-6}
                                                                                      & \begin{tabular}[c]{@{}c@{}}Video\\ on drones\end{tabular}                   & \begin{tabular}[c]{@{}c@{}}Autonomous\\ drones\end{tabular}        & \begin{tabular}[c]{@{}c@{}}Portable\\ edge computers\end{tabular}        & \begin{tabular}[c]{@{}c@{}}Edge\\ only\end{tabular}                         & \begin{tabular}[c]{@{}c@{}}Bandwidth\\ saving\end{tabular}                               \\ \hline
\begin{tabular}[c]{@{}c@{}}Virtual\\ Reality\end{tabular}                             & MUVR                                                                        & Smartphones                                                        & \begin{tabular}[c]{@{}c@{}}Individual\\ households\end{tabular}          & \begin{tabular}[c]{@{}c@{}}Edge\\ only\end{tabular}                         & \begin{tabular}[c]{@{}c@{}}Resource utilization\\ efficiency\\ optimization\end{tabular} \\ \hline
\end{tabular}
\end{table*}

\textit{1) Service management module:} 
This type of module is designed for the management of data views. It provides interfaces to update the data view, as well as relevant programs for data processing.

\textit{2) Job management module:}
This type of module is responsible for the scheduling, monitoring, and evaluation of the task executions. When the local computing resources are insufficient, the module can look into the node list and the data view\nop{ (of the Firework manager)} and make resource re-scheduling at the same sub-service layer. When the sub-service is running, the module can also provide necessary monitoring information and give feedback to other upstream and downstream nodes for flow layer scheduling.

\textit{3) Actuator management module:}
This type of module is mainly responsible for managing all hardware resources and hosting the execution processes of different tasks. With the help of this module, the device, running environment and the upper layer functions could be decoupled, so that the nodes of Firework system are not limited to a certain type of devices, and the data processing environment is not limited to a certain type of computing platform.

\nop{
Firework manager has two sub-modules, and Firework computing node has three sub-modules. For Firework manager, it includes service management module that provides the registration and viewing capabilities of the view and the status of the associated Firework compute nodes, and job management module, which is responsible for scheduling the service. Firework computing node refers to a node that provides computing resources, which may have its own database and provide a corresponding view of the data but may also simply provide computing resources. 
}

\subsection{Other Edge Computing Systems}

The edge systems introduced above depict some typical and basic innovations on the exploitation of edge analytics for highly-responsive services. The previous systems are used in general cases, which lay the foundation of further development. We highlight that, in addition to these efforts, there are many other edge systems tuned for a series of different application scenarios.In Table~\ref{tab:otherEdgeSystems}, we briefly summarize the general and application-specific edge systems, which leverage different kinds of edge nodes to serve diverse end devices using either a hybrid or edge only computation architecture.

Compared to both on-board computation and cloud-based computation, edge computing can provide more effective data analytics with lower latency for the moving vehicles~\cite{zhang2018openvdap}. In~\cite{zhang2018openvdap}, an open full-stack edge computing-based platform OpenVDAP is proposed for the data analytics of connected and autonomous vehicles (CAVs). OpenVDAP proposes systematic mechanisms, including varied wireless interfaces, to utilize the heterogeneous computation resources of nearby CAVs, edge nodes, and the cloud. For optimal utilization of the resources, a dynamic scheduling interface is also provided to sense the status of available resources and to offload divided tasks in a distributed way for computation efficiency. SafeShareRide is an edge-based attack detection system addressing in-vehicle security for ridesharing services~\cite{liu2018safeshareride}. Its three detection stages leverage both the smartphones of drivers and passengers as edge computing platform to collect multimedia information in vehicles. Specifically, speech recognition and driving behavior detection stages are first carried out independently to capture in-vehicle danger, and video capture and uploading stage is activated when abnormal keywords or dangerous behaviors are detected to collect videos for cloud-based analysis. By using such an edge-cloud collaborative architecture, SafeShareRide can accurately detect attacks in-vehicle with low bandwidth demand.

Another scenario that edge computing can play an important role is the IoT devices management in the smart home environment. Wherein, the privacy issue of the wide range of home-devices is a popular topic. In~\cite{trimananda2018vigilia}, the Vigilia system is proposed to harden smart home systems by restricting the network access of devices. A default access deny policy and an API-granularity device access mechanism for applications are adopted to enforce access at the network level. Run time checking implemented in the routers only permits those declared communications, thus helping users secure their home-devices. Similarly, the HomePad system in~\cite{zavalyshyn2018homepad} also proposes to execute IoT applications at the edge and introduces a privacy-aware hub to mitigate security concerns. Homepad allows users to specify privacy policy to regulate how applications access and process their data. Through enforcing applications to use explicit information flow, Homepad can use Prolog rules to verify whether applications have the ability to violate the defined privacy policy at install time.

Edge computing has also been widely used in the analysis of video stream. LAVEA is an edge-based system built for latency-aware video analytics nearby the end users~\cite{yi2017lavea}. In order to minimize the response time, LAVEA formulates an optimization problem to determine which part of tasks to be offloaded to the edge computer and uses a task queue prioritizer to minimize the makespan. It also proposes several task placement schemes to enable the collaboration of nearby edge nodes, which can further reduce the overall task completion time. VideoEdge is a system that provides the most promising video analytics implementation across a hierarchy of clusters in the city environment~\cite{hung2018videoedge}. A 3-tier computation architecture is considered with deployed cameras and private clusters as the edge and remote server as the cloud. The hierarchical edge architecture is also adopted in~\cite{tong2016hierarchical} and is believed to be promising in processing live video stream at scale. Technically, VideoEdge searches thousands of combinations of computer vision components implementation, knobs, and placement and finds a configuration to balance the accuracy and resource demands using an efficient heuristic. In~\cite{wang2018bandwidth}, a video analytics system for autonomous drones is proposed, where edge computing is introduced to save the bandwidth. Portable edge computers are required here to support dynamic transportation during a mission. Totally four different video transmission strategies are presented to build an adaptive and efficient computer vision pipeline. In addition to the analytics work (e.g., object recognition), the edge nodes also train filters for the drones to avoid the uploading of the uninteresting video frames.

In order to provide flexible virtual reality (VR) on untethered smartphones, edge computing can be useful to transport the heavy workload from smartphones to their nearby edge cloud~\cite{li2018muvr}. However, the rendering task of the panoramic VR frames (i.e., 2GB per second) will also saturate the individual households as common edge in the house. In~\cite{li2018muvr}, the MUVR system is designed to support multi-user VR with efficient bandwidth and computation resources utilization. MUVR is built on a basic observation that the VR frames being rendered and transmitted to different users are highly redundant. For computation efficiency, MUVR maintains a two-level hierarchical cache for invariant background at the edge and the user end to reuse frames whenever necessary. Meanwhile, MUVR transmits a part of all frames in full and delivers the distinct portion for the rest frames to further reduce the transmission costs.

\section{Open Source Edge Computing Projects}
\label{sec:open-source-projects}

Besides the designed edge computing systems for specific purposes, some open source edge computing projects have also been launched recently. The Linux Foundation published two projects, EdgeX Foundry in 2017 and Akraino Edge Statck~\cite{akrainoEdge2018} in 2018. The Open Network Foundation(ONF) launched a project namely CORD (Central Office Re-architected as a Datacenter)~\cite{CORD2018}. The Apache Software Foundation published Apache Edgent. Microsoft published Azure IoT Edge in 2017 and announced it as an open source in 2018.

Among them, CORD and Akraino Edge Stack focus on providing edge cloud services; EdgeX Foundry and Apache Edgent focus on IoT and aim to solve problems which bring difficulties to practical applications of edge computing in IoT; Azure IoT Edge provides hybrid cloud-edge analytics, which helps to migrate cloud solutions to IoT devices.

\subsection{CORD}

\begin{figure}[!tb]
\begin{center}
\includegraphics[width=0.5\textwidth]{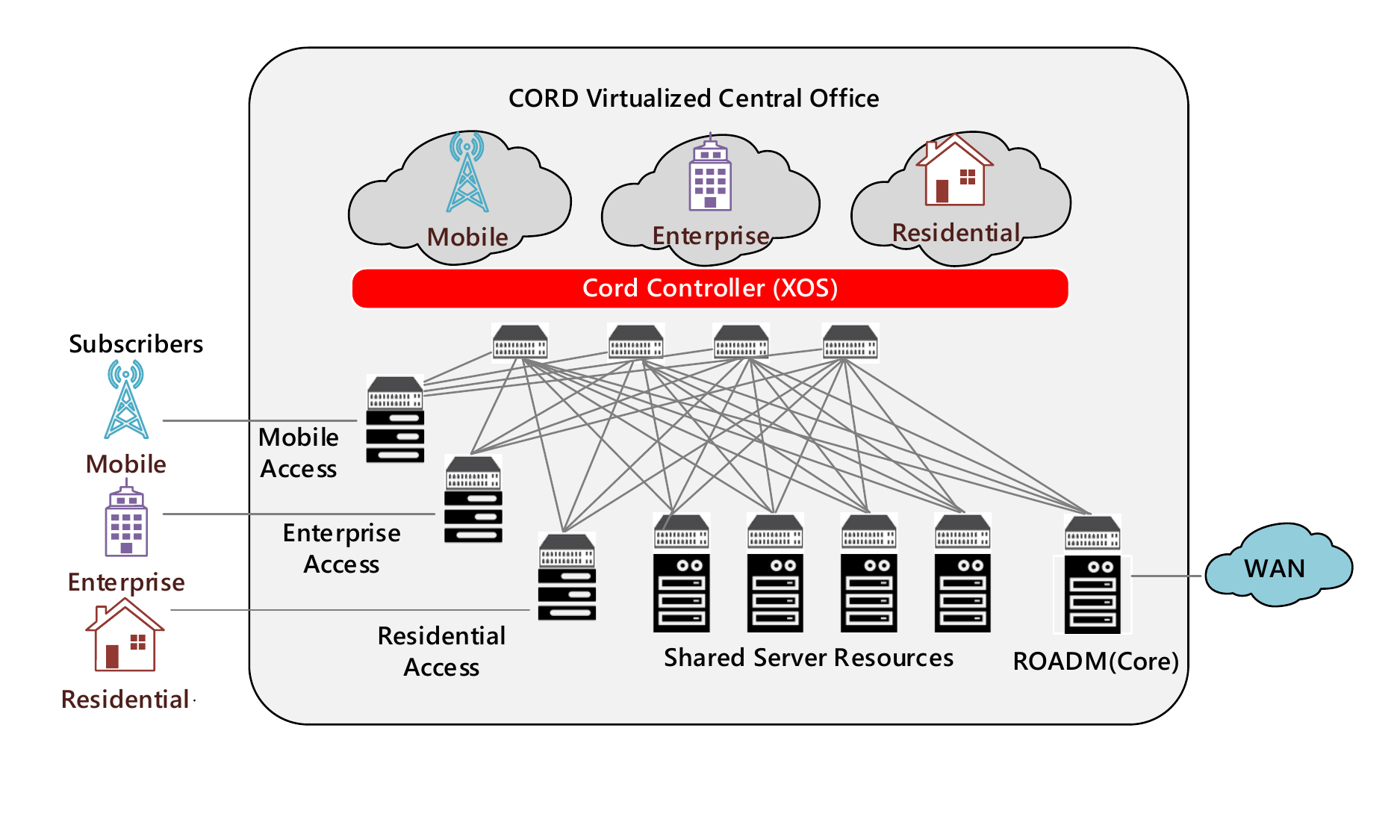}
\caption{Hardware Architecture of CORD.}\label{fig:CORD}
%\vspace{-0.20in}
\end{center}
\end{figure} 

CORD is an open source project of ONF initiated by AT\&T and is designed for network operators. Current network infrastructure is built with closed proprietary integrated systems provided by network equipment providers. Due to the closed property, the network capability cannot scale up and down dynamically. And the lack of flexibility results in inefficient utilization of the computing and networking resources. CORD plans to reconstruct the edge network infrastructure to build datacenters with SDN~\cite{nunes2014survey}, NFV~\cite{hawilo2014nfv} and Cloud technologies. It attempts to slice the computing, storage and network resources so that these datacenters can act as clouds at the edge, providing agile services for end users.

\begin{figure}[!tb]
\begin{center}
\includegraphics[width=0.5\textwidth]{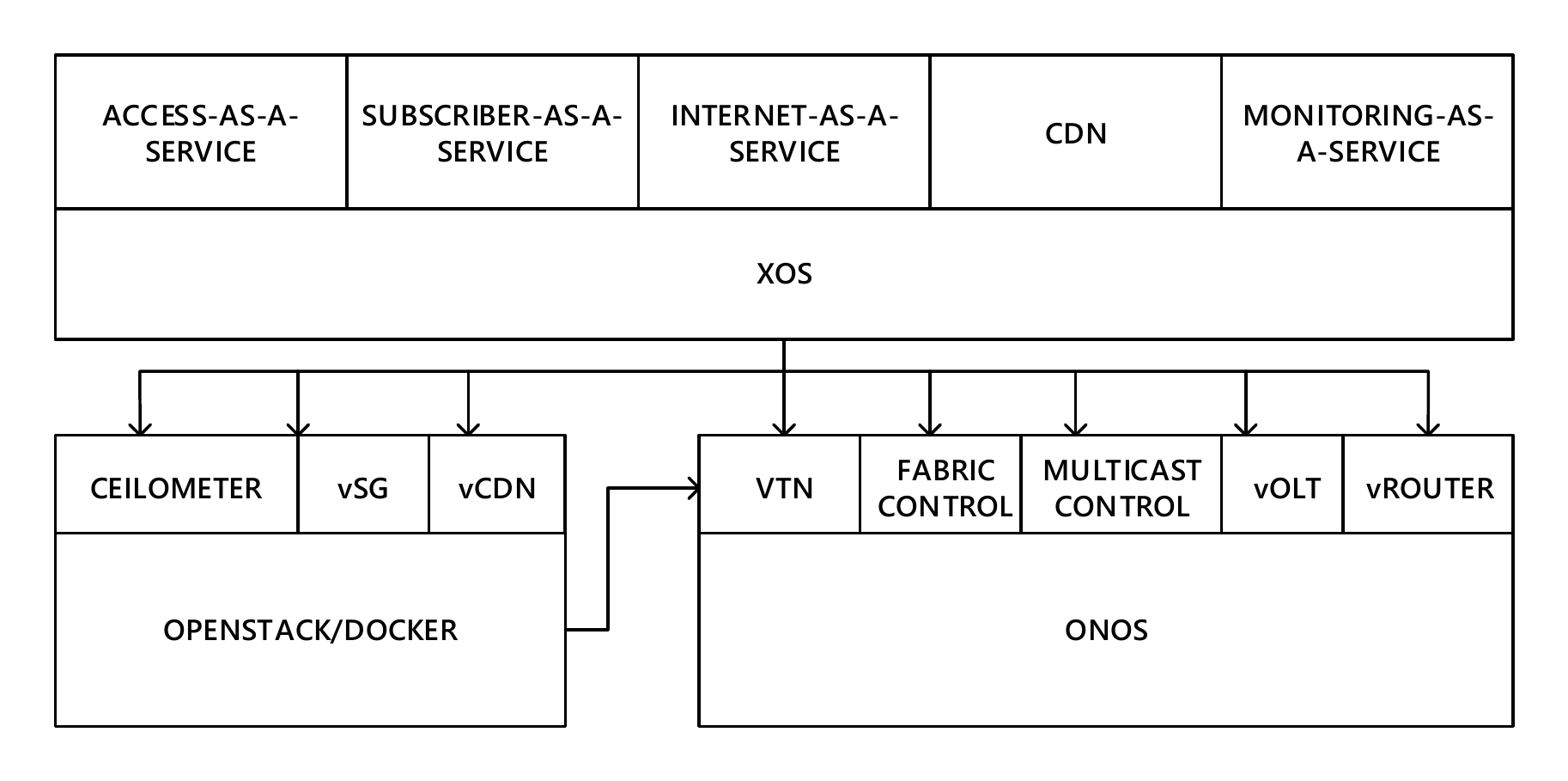}
\caption{Software Architecture of CORD.}\label{fig:CORD_SW}
%\vspace{-0.20in}
\end{center}
\end{figure}

CORD is an integrated system built from commodity hardware and open source software. Fig.~\ref{fig:CORD} shows the hardware architecture of CORD~\cite{CORD2018}. It uses commodity servers that are interconnected by a Fabric of White-box switches. White-box switch~\cite{Manggala2016Performance} is  a  component of SDN switch, which is responsible to regulate the flow of data according to SDN controller. These commodity servers provide computing, storage resources, and the fabric of switches are used to build the network. This switching fabric is organized to a Spine-Leaf topology \cite{Okafor2017Leveraging}, a kind of flat network topology structure which adds a horizontal network structure parallel to the trunk longitudinal network structure, and then adds corresponding switching network on the horizontal structure. Comparing to traditional three-tier network topology, it can provide scalable throughput for greater East-to-West network traffic, that is, traffic coming from network diagram drawings that usually depict local area network (LAN) traffic horizontally. In addition, specialized access hardware is required to connect subscribers. The subscribers can be divided into three categories for different use cases, mobile subscribers, enterprise subscribers and residential subscribers. Each category demands different access hardware due to different access technologies. In terms of software, Fig.~\ref{fig:CORD_SW} shows the software architecture of CORD~\cite{CORD2018}. Based on the servers and the fabric of switches, OpenStack provides with IaaS capability for CORD, it manages the compute, storage and networking resources as well as creating virtual machines and virtual networks. Docker is used to run services in containers for isolation. ONOS(Open Network Operating System) is a network operating system which is used to manage network components like the switching fabric and provide communication services to end-users. XOS provides a control plane to assemble and compose services. Other software projects provide component capabilities, for example, vRouter(Virtual Router) provides with virtual routing functionality.

The edge of the operator network is a sweet spot for edge computing because it connects customers with operators and is close to customers' applications as data sources. CORD takes edge computing into consideration and moves to support edge computing as a platform to provide edge cloud services (from the released version $4.1$). CORD can be deployed into three solution, M-CORD (Mobile CORD), R-CORD (Residential CORD) and E-CORD (Enterprise CORD) for different use cases. M-CORD focuses on mobile network, especially 5G network, and it plans to disaggregate and virtualize cellular network functions to enable services be created and scaled dynamically. This agility helps to provide multi-access edge services for mobile applications. For those use cases like driverless cars or drones, users can rent the edge service to run their edge applications. Similarly, R-CORD and E-CORD are designed to be agile service delivery platforms but for different users, residential and enterprise users relatively. 

So far, the deployment of CORD is still under test among network operators, and more researches are needed to combine CORD with various edge applications.

\subsection{Akraino Edge Stack}

Akraino Edge Stack, initiated by AT\&T and now hosted by Linux Foundation, is a project to develop a holistic solution for edge infrastructure so as to support high-availability edge cloud services~\cite{akrainoEdge2018}. An open source software stack, as the software part of this solution, is developed for network carrier to facilitate optimal networking and workload orchestration for underlying infrastructure in order to meet the need of edge computing such as low latency, high performance, high availability, scalability and so on.

\begin{figure}[!tb]
\begin{center}
\includegraphics[width=0.5\textwidth]{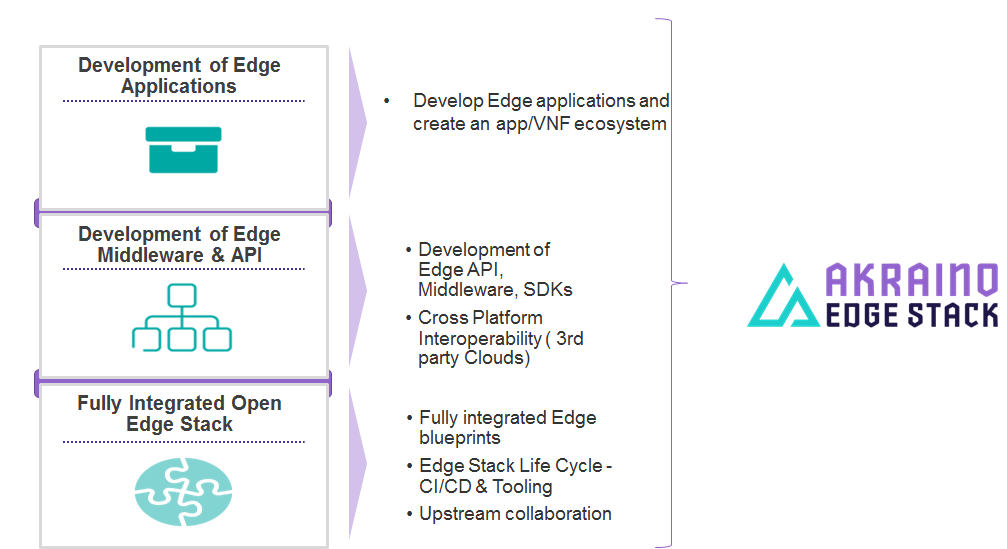}
\caption{Akraino Edge Stack's Scope.}\label{fig:Akraino}
%\vspace{-0.20in}
\end{center}
\end{figure}

To provide a holistic solution, Akraino Edge Stack has a wide scope from infrastructure layer to application layer. Fig.~\ref{fig:Akraino}~\cite{akrainoEdge2018} shows the scope with three layers. In the application layer, Akraino Edge Stack wants to create a virutal network function (VNF) ecosystem and calls for edge applications. The second layer consists of middleware which supports applications in the top layer. In this layer, Akraino Edge Stack plans to develop Edge API and framework for interoperability with 3rd party Edge projects such as EdgeX Foundry. At the bottom layer, Akraino Edge Stack intends to develop an open source software stack for the edge infrastructure in collaboration with upstream communities. It interfaces with and maximizes the use of existing open source projects such as Kubernetes, OpenStack and so on. Akraino Edge Stack provides different edge use cases with blueprints, which are declarative configurations of entire stack including hardware, software, point of delivery, etc~\cite{akrainoEdge2018}. The application domains of these blueprints start from Telco industry and are expected to applied in more domains like Enterprise and industrial IoT. Now Akraino Edge Stack has put forward several blueprints such as Micro-MEC and Edge Media Processing. Micro-MEC intends to develop a new service infrastructure for smart cities, which enables developing services for smart city and has high data capacity for citizens. Edge Media Processing intends to develop a network cloud to enable real-time media processing and edge media AI analytics with low latency.

As an emerging project, Akraino Edge Stack has been taken to execution since August 2018. Thus more researches need to be done with the development of this project.

\subsection{EdgeX Foundry}
EdgeX Foundry is a standardized interoperability framework for IoT edge computing, whose sweet spots are edge nodes such as gateways, hubs, and routers~\cite{edgexF2018}. It can connect with various sensors and devices via different protocols, manage them and collect data from them, and export the data to a local application at the edge or the cloud for further processing. EdgeX is designed to be agnostic to hardware, CPU, operating system, and application environment. It can run natively or run in docker containers.

\begin{figure*}[!tb]
\begin{center}
\includegraphics[width=0.65\textwidth]{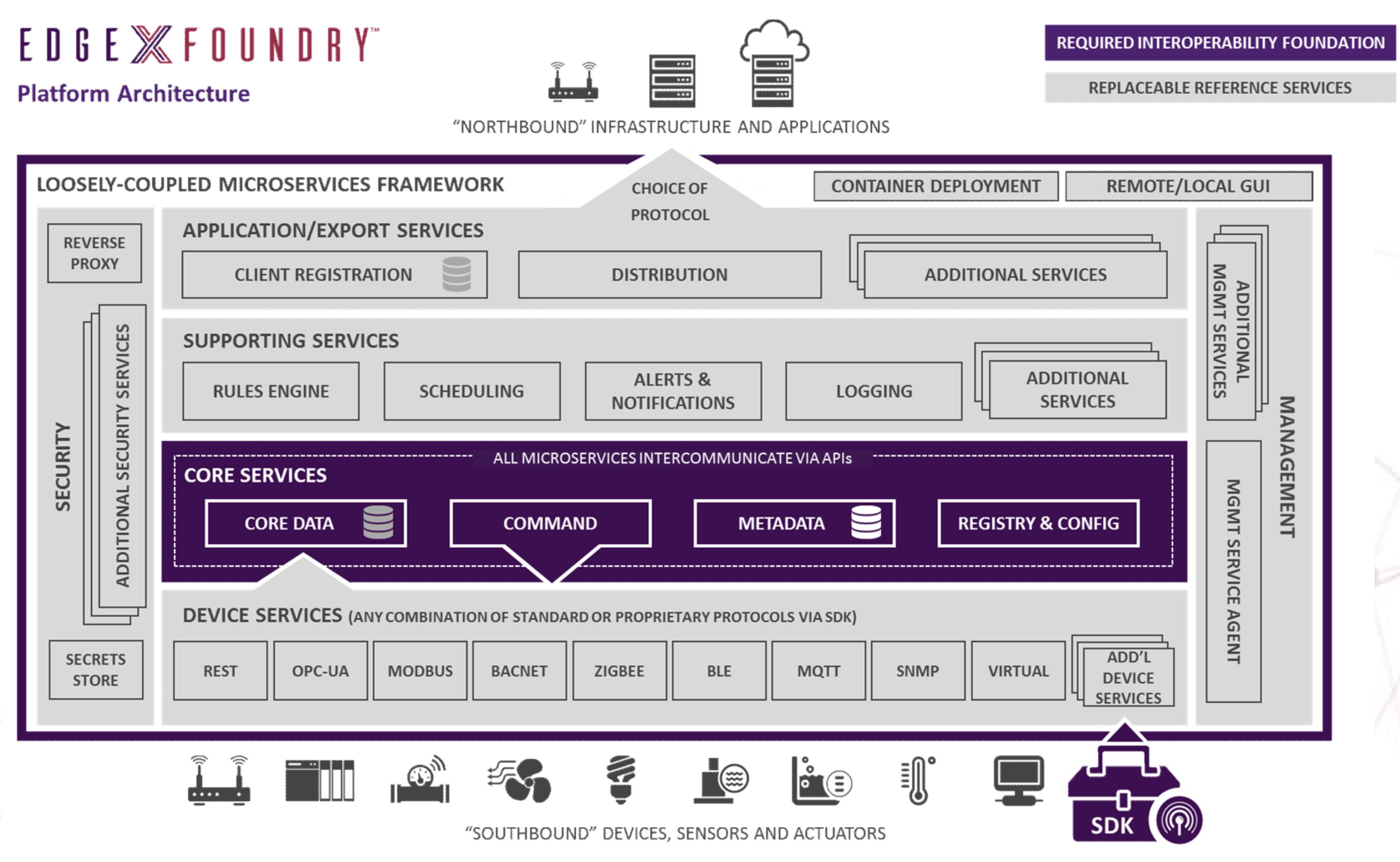}
\caption{Architecture of EdgeX Foundry.}\label{fig:EdgeX}
%\vspace{-0.20in}
\end{center}
\end{figure*}

Fig.~\ref{fig:EdgeX}~\cite{edgexF2018} shows the architecture of EdgeX Foundry. ``South side'' at the bottom of the figure includes all IoT objects, and the edge of the network that communicates directly with those devices, sensors, actuators and other IoT objects to collect the data from them. Relatively, ``north side'' at the top of the figure includes the Cloud (or Enterprise system) where data are collected, stored, aggregated, analyzed, and turned into information, and the part of the network that communicates with the Cloud. EdgeX Foundry connects these two sides regardless of the differences of hardware, software and network. EdgeX tries to unify the manipulation method of the IoT objects from the south side to a common API, so that those objects can be manipulated in the same way by the applications of north side.

EdgeX uses a Device Profile to describe a south side object. A Device Profile defines the type of the object, the format of data that the object provides, the format of data to be stored in EdgeX and the commands used to manipulate this object. Each Device Profile involves a Device Service, which is a service that converts the format of the data, and translates the commands into instructions that IoT objects know how to execute. EdgeX provides SDK for developers to create Device Services, so that it can support for any combination of device interfaces and protocols by programming.

EdgeX consists of a collection of microservices, which allows services to scale up and down based on device capability. These microservices can be grouped into four service layers and two underlying augmenting system services, as depicted in Fig.~\ref{fig:EdgeX}. The four service layers include Device Services Layer, Core Services Layer, Supporting Services Layer and Export Services Layer, respectively; the two underlying augmenting system services are System Management and Security, respectively. Each of the six layers consists of several components and all components use a common Restful API for configuration.

\textit{1) Device Services Layer:}
This layer consists of Device Services. According to the Device Profiles, Device Service Layer converts the format of the data, sends them to Core Services Layer, and translates the command requests from the Core Services Layer.

\textit{2) Core Services Layer:}
This layer consists of four components: Core Data, Command, Metadata, and Registry \& Configuration. Core Data is a persistence repository as well as a management service. It stores and manages the data collected from the south side objects. Command is a service to offer the API for command requests from the north side to Device Services. Metadata is a repository and management service for metadata about IoT objects. For example, the Device Profiles are uploaded and stored in Metadata. Registry \& Configuration provides centralized management of configuration and operating parameters for other microservices. 

\textit{3) Supporting Services Layer:}
This layer is designed to provide edge analytics and intelligence~\cite{hawilo2014nfv}. Now the Rules Engine, Alerting and Notification, Scheduling and Logging microservices are implemented. A target range of data can be set to trigger a specific device actuation as a rule and Rules Engine helps realize the rule by monitoring the incoming data. Alerting and Notifications can send notifications or alerts to another system or person by email, REST callback or other methods when an urgent actuation or a service malfunction happens. The scheduling module can set up a timer to regularly clean up the stale data. Logging is used to record the running information of EdgeX.

\textit{4) Export Services Layer:}
This layer connects EdgeX with North Side and consists of Client Registration and Export Distribution. Client Registration enables clients like a specific cloud or a local application to register as recipients of data from Core Data. Export Distribution distributes the data to the Clients registered in Client Registration. 

\textit{5) System Management and Security:}
System Management provides management operations including installation, upgrade, starting, stopping and monitoring, as EdgeX is scalable and can be deployed dynamically. Security is designed to protect the data and command of IoT objects connected with EdgeX Foundry.

EdgeX is designed for the user cases dealing with multitudes of sensors or devices, such as automated factories, machinery systems and lots of other cases in IoT. Now EdgeX Foundry is in the rapid upgrading phase, and more features will be added in future releases. An EdgeX UI is in development as a web-based interface to add and manage the devices.

\subsection{Apache Edgent}
Apache Edgent, which was known as Apache Quarks previously, is an Apache Incubator project at present. It is an open source programming model for lightweight runtime data analytics, used in small devices such as routers and gateways at the edge. Apache Edgent focuses on data analytics at the edge, aiming to accelerate the development of data analysis. 

\begin{figure}[!tb]
\begin{center}
\includegraphics[width=0.5\textwidth]{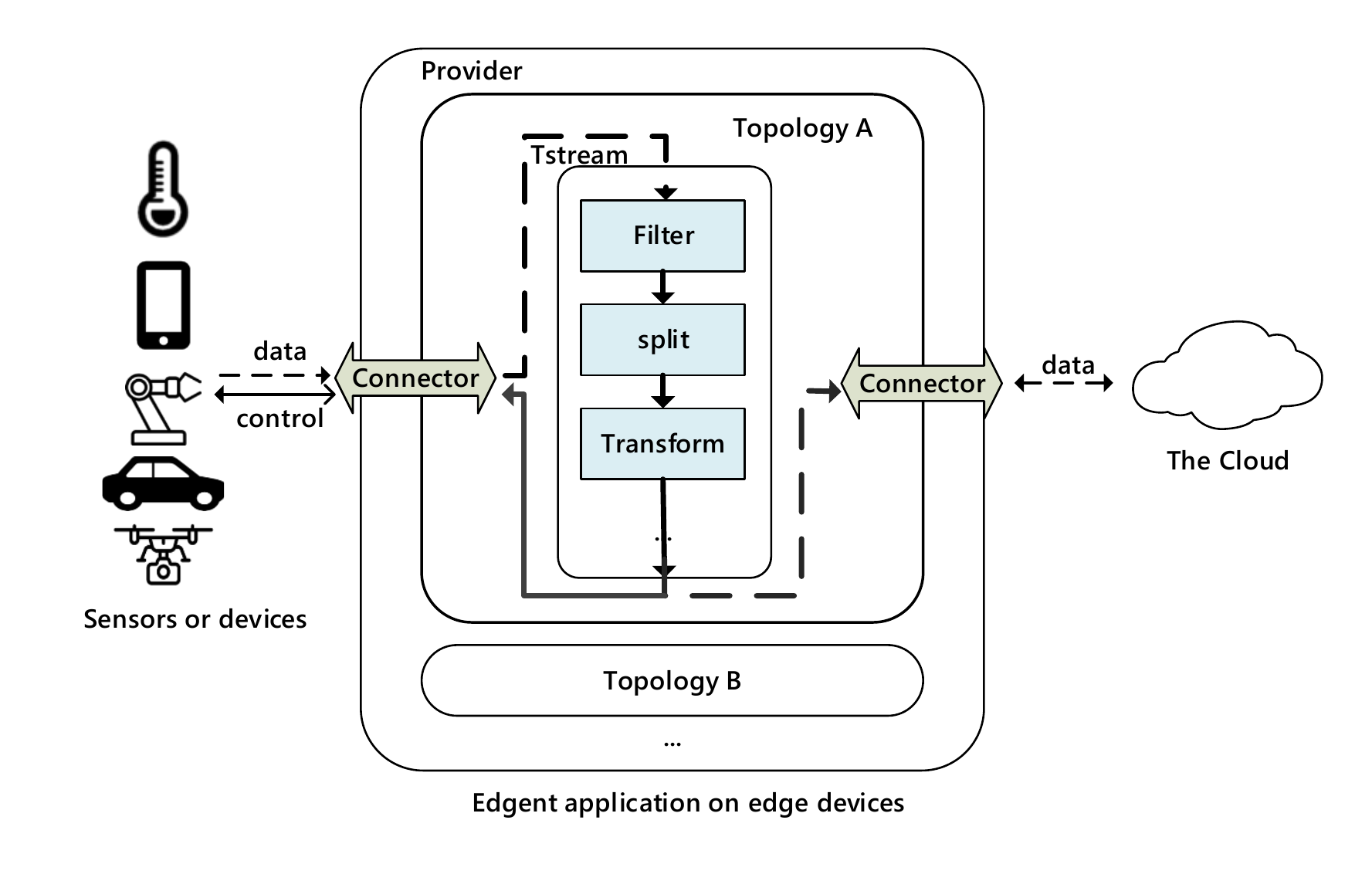}
\caption{Model of the Edgent Applications.}\label{fig:Edgent}
%\vspace{-0.20in}
\end{center}
\end{figure}

As a programming model, Edgent provides API to build edge applications. Fig.~\ref{fig:Edgent} illustrates the model of the Edgent applications. Edgent uses a topology as a graph to represent the processing transformation of streams of data which are abstracted to a Tstream class. A connector is used to get streams of data from external entities such as sensors and devices in physical world, or to send streams of data to back-end systems like a cloud. The primary API of Edgent is responsible for data analysis. The streams of data can be filtered, split, transformed or processed by other operations in a topology. Edgent uses a provider to act as a factory to create and execute topologies. To build an Edgent application, users should firstly get a provider, then create a topology and add the processing flow to deal with the streams of data, and finally submit the topology. The deployment environments of Edgent are Java 8, Java 7 and Android. 

Edgent provides APIs for sending data to back-end systems and now supports MQTT, IBM Watson IoT Platform, Apache Kafka and custom message hubs. Edgent applications analyze the data from sensors and devices, and send the essential data to the back-end system for further analysis. For IoT use cases, Edgent helps reduce the cost of transmitting data and provide local feedback. 

Edgent is suitable for use cases in IoT such as intelligent transportation, automated factories and so on. In addition, the data in Edgent applications are not limited to sensor readings, they can also be files or logs. Therefore, Edgent can be applied to other use cases. For example, it can perform local data analysis when embedded in application servers, where it can analyze error logs without impacting network traffic~\cite{apacheEdgent2018}.

\subsection{Azure IoT Edge}\label{subsec:AzureIoT}

\begin{figure}[!tb]
\begin{center}
\includegraphics[width=0.5\textwidth]{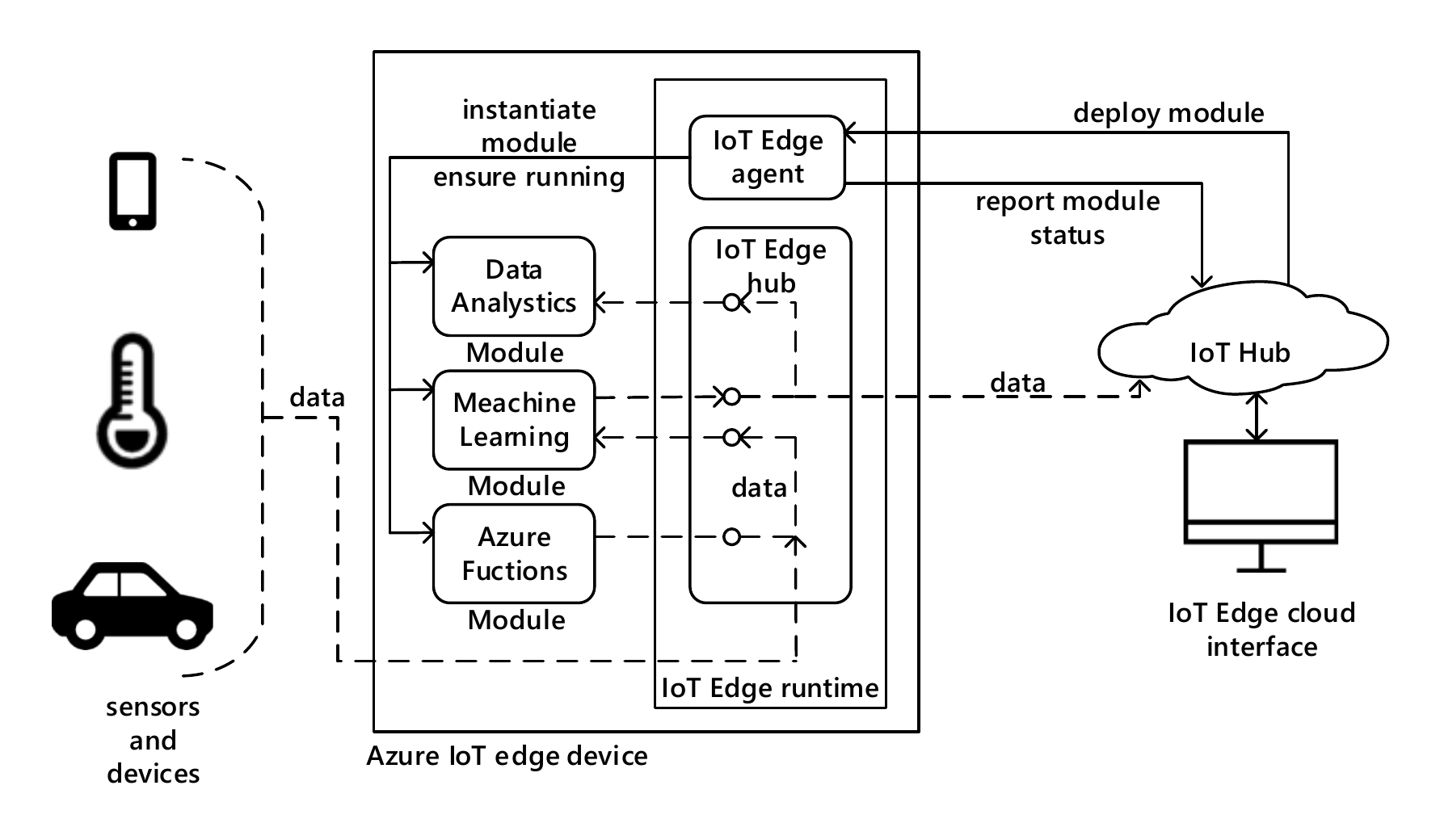}
\caption{Diagram of Azure IoT Edge.}\label{fig:AzureIoT}
%\vspace{-0.20in}
\end{center}
\end{figure}

Azure IoT Edge, provided by Microsoft Azure as a cloud service provider, tries to move cloud analytics to edge devices. These edge devices can be routers, gateways or other devices which can provide computing resources. The programming model of Azure IoT Edge is the same as that of other Azure IoT services~\cite{azureIoT2018} in the cloud, which enables user to move their existing applications from Azure to the edge devices for lower latency. The convenience simplifies the development of edge applications. In addition, Azure services like Azure Functions, Azure Machine Learning and Azure Stream Analytics can be used to deploy complex tasks on the edge devices such as machine learning, image recognition and other tasks about artificial intelligence.

Azure IoT Edge consists of three components: IoT Edge modules, IoT Edge runtime and a cloud-based interface as depicted in Fig.~\ref{fig:AzureIoT}. The first two components run on edge devices, and the last one is an interface in the cloud. IoT Edge modules are containerized instances running the customer code or Azure services. IoT Edge runtime manages these modules. The cloud-based interface is used to monitor and manage the former two components, in other words, monitor and manage the edge devices.

IoT Edge modules are the places that run specific applications as the units of execution. A module image is a docker image containing the user code. A module instance, as a docker container, is a unit of computation running the module image. If the resources at the edge devices are sufficient, these modules can run the same Azure services or custom applications as in the cloud because of the same programming model. In addition, these modules can be deployed dynamically as Azure IoT Edge is scalable.

IoT Edge runtime acts as a manager on the edge devices. It consists of two modules: IoT Edge hub and IoT Edge agent. IoT Edge hub acts as a local proxy for IoT Hub which is a managed service and a central message hub in the cloud. As a message broker, IoT Edge hub helps modules communicate with each other, and transport data to IoT Hub. IoT Edge agent is used to deploy and monitor the IoT Edge modules. It receives the deployment information about modules from IoT Hub, instantiates these modules, and ensures they are running, for example, restarts the crashed modules. In addition, it reports the status of the modules to the IoT hub.

IoT Edge cloud interface is provided for device management. By this interface, users can create edge applications, then send these applications to the device and monitor the running status of the device. This monitoring function is useful for use cases with massive devices, where users can deploy applications to devices on a large scale and monitor these devices.

A simple deployment procedure for applications is that: users choose a Azure service or write their own code as an application, build it as an IoT Edge module image, and deploy this module image to the edge device with the help of the IoT Edge interface. Then the IoT Edge receives the deployment information, pulls the module image, and instantiates the module instance.

Azure IoT Edge has wide application areas. Now it has application cases on intelligent manufacturing, irrigation system, drone management system and so on. It is worth noting that Azure IoT Edge is open-source but the Azure services like Azure Functions, Azure Machine Learning and Azure Stream are charged.

\subsection{Comparative Study}

We summarize the features of the above open source edge computing systems in Table~\ref{tab:characterEdge}. Then we compare them from different aspects in Table~\ref{tab:characterEdge}, including the main purpose of the systems,application  area,  deployment, target  user,  virtualization  technology,  system  characteristic, limitations, scalability and mobility. We believe such comparisons give better understandings of current open source edge computing systems. 

\begin{table*}[htb]
\centering
\caption{Comparison of Open Edge System Characteristics}
\label{tab:characterEdge}
\begin{tabular}{cccccc}
\hline
\textbf{Aspect}                                                               & \textbf{EdgeX Foundry}                                                                          & \begin{tabular}[c]{@{}c@{}}\textbf{Azure IoT}\\ \textbf{Edge}\end{tabular}                        & \textbf{Apache Edgent}                                                                                      & \textbf{CORD}                                                                                                                       & \begin{tabular}[c]{@{}c@{}}\textbf{Akraino}\\ \textbf{Edge Stack}\end{tabular}
                                  \\ \hline
\begin{tabular}[c]{@{}c@{}}User access\\ interface\end{tabular} & \begin{tabular}[c]{@{}c@{}}Restful API\\ or EdgeX UI\end{tabular} & \begin{tabular}[c]{@{}c@{}}Web service,\\ Command-line\end{tabular}   & API           & \begin{tabular}[c]{@{}c@{}}API or\\ XOS-GUI\end{tabular}       & N/A                                                                                      \\ \hline
OS support                                                      & Various OS                                                        & Various OS                                                            & Various OS    & Ubuntu                                                         & Linux                                                        \\ \hline
\begin{tabular}[c]{@{}c@{}}Programming\\ framework\end{tabular} & Not provided                                                      & \begin{tabular}[c]{@{}c@{}}Java, .NET, C,\\ Python, etc.\end{tabular} & Java          & \begin{tabular}[c]{@{}c@{}}Shell script,\\ Python\end{tabular} &N/A
\\ \hline
Main purpose                                                         & \begin{tabular}[c]{@{}c@{}}Provide with\\ Interoperability\\ for IoT edge\end{tabular} & \begin{tabular}[c]{@{}c@{}}Support hybrid\\ cloud-edge\\ analytics\end{tabular} & \begin{tabular}[c]{@{}c@{}}Accelerate the \\ development\\ process of\\ data analysis\end{tabular} & \begin{tabular}[c]{@{}c@{}}Transform edge of\\ the operator network\\ into agile service\\ delivery platforms\end{tabular} & \begin{tabular}[c]{@{}c@{}}Support edge\\ clouds with an\\ open source\\ software stack\end{tabular} \\ \hline
Application area                                                     & IoT                                                                                    & Unrestricted                                                                    & IoT                                                                                                & Unrestricted                                                                                                               & Unrestricted                                                  \\ \hline
Deployment                                                      & Dynamic                                                           & Dynamic                                                               & Static        & Dynamic                                                        & Dynamic                                                                 \\ \hline
Target user                                                          & General users                                                                          & General users                                                                   & General users                                                                                      & Network operators                                                                                                          & \begin{tabular}[c]{@{}c@{}}Network\\ operators\end{tabular}                                          \\ \hline
\begin{tabular}[c]{@{}c@{}}Virtualization \\ technology\end{tabular} & Container                                                                              & Container                                                                       & JVM                                                                                                & \begin{tabular}[c]{@{}c@{}}Virtual Machine\\ and Container\end{tabular}                                                    & \begin{tabular}[c]{@{}c@{}}Virtual Machine\\ and Container\end{tabular}                              \\ \hline
\begin{tabular}[c]{@{}c@{}}System\\ characteristics\end{tabular}     & \begin{tabular}[c]{@{}c@{}}A common API for\\ device management\end{tabular}           & \begin{tabular}[c]{@{}c@{}}Powerful\\ Azure services\end{tabular}               & \begin{tabular}[c]{@{}c@{}}APIs for\\ data analytics\end{tabular}                                  & \begin{tabular}[c]{@{}c@{}}Widespread\\ edge clouds\end{tabular}                                                           & \begin{tabular}[c]{@{}c@{}}Widespread\\ edge clouds\end{tabular}                                     \\ \hline
Limitation                                                           & \begin{tabular}[c]{@{}c@{}}Lack of\\ programable\\ interface\end{tabular}              & \begin{tabular}[c]{@{}c@{}}Azure Services\\ is chargeable\end{tabular}          & \begin{tabular}[c]{@{}c@{}}Limited to\\ data analytics\end{tabular}                                & \begin{tabular}[c]{@{}c@{}}Unable to\\ be offline\end{tabular}                                                             & \begin{tabular}[c]{@{}c@{}}Unable to\\ be offline\end{tabular} 
    \\ \hline
\begin{tabular}[c]{@{}c@{}}Scalability\end{tabular}     & \begin{tabular}[c]{@{}c@{}}Scalable\end{tabular}           & \begin{tabular}[c]{@{}c@{}}Scalable\end{tabular}               & \begin{tabular}[c]{@{}c@{}}Not scalable\end{tabular}                                  & \begin{tabular}[c]{@{}c@{}}Scalable\end{tabular}                                                           & \begin{tabular}[c]{@{}c@{}}Scalable\end{tabular}
    \\ \hline
\begin{tabular}[c]{@{}c@{}}Mobility\end{tabular}     & \begin{tabular}[c]{@{}c@{}}Not support\end{tabular}           & \begin{tabular}[c]{@{}c@{}}Not support\end{tabular}               & \begin{tabular}[c]{@{}c@{}}Not support\end{tabular}                                  & \begin{tabular}[c]{@{}c@{}}Support\end{tabular}                                                           & \begin{tabular}[c]{@{}c@{}}Support\end{tabular}
\\ \hline
\end{tabular}
\end{table*}

\subsubsection{Main purpose}
The main purpose shows the target problem that a system tries to fix, and it is a key factor for us to choose a suitable system to run edge applications. As an interoperability framework, EdgeX Foundry aims to communicate with any sensor or device in IoT. This ability is necessary for edge applications with data from various sensors and devices. Azure IoT Edge offers an efficient solution to move the existing applications from cloud to edge, and to develop edge applications in the same way with the cloud applications. Apache Edgent helps to accelerate the development process of data analysis in IoT use cases. CORD aims to reconstruct current edge network infrastructure to build datacenters so as to provide agile network services for end-user customers. From the view of edge computing, CORD provides with multi-access edge services. Akraino Edge Stack provides an open source software stack to support high-availability edge clouds.

\subsubsection{Application area}
EdgeX Foundry and Apache Edgent both focus on IoT edge, and EdgeX Foundry is geared toward communication with various sensors and devices, while Edgent is geared toward data analysis. They are suitable for intelligent manufacturing, intelligent transportation and smart city where various sensors and devices generate data all the time. Azure IoT Edge can be thought as the expansion of Azure Cloud. It has an extensive application area but depends on the compute resources of edge devices. Besides, it is very convenient to deploy edge applications about artificial intelligence such as machine learning and image recognition to Azure IoT Edge with the help of Azure services. CORD and Akraino Edge Stack support edge cloud services, which have no restriction on application area. If the edge devices of users don't have sufficient computing capability, these two systems are suitable for users to run resource-intensive and interactive applications in connection with operator network.

\subsubsection{Deployment}
As for the deployment requirements, EdgeX Foundry, Apache Edgent and Azure IoT Edge are deployed in edge devices such as routers, gateways, switchers and so on. Users can deploy EdgeX Foundry by themselves, add or reduce microservices dynamically, and run their own edge applications. Differently, users need the help of cloud-based interface to deploy Azure IoT Edge and develop their edge applications. CORD and Akraino Edge Stack are designed for network operators, who need fabric switches, access devices, network cards and other related hardware apart from compute machines. Customers have no need to think about the hardware requirements and management process of the hardware, but to rent the services provided by the network operators like renting a cloud service instead of managing a physical server.

\subsubsection{Target user}
Though these open source systems focus on edge computing, their target users are not the same. EdgeX Foundry, Azure IoT Edge and Apache Edgent have no restriction on target users. Therefore, every developer can deploy them into local edge devices like gateways, routers and hubs. Differently, CORD and Akraino Edge Stack are created for network operators because they focus on edge infrastructure.

\subsubsection{Virtualization technology}
Virtualization technologies are widely used nowadays. Virtual machine technology can provide better management and higher utilization of resources, stability, scalability and other advantages. Container technology can provide services with isolation and agility but with negligible overhead, which can be used in edge devices~\cite{morabito2017virtualization}. Using OpenStack and Docker as software components, CORD and Akraino Edge Stack use both of these two technologies to support edge cloud. Different edge devices may have different hardware and software environment. For those edge systems which are deployed on edge devices, container is a good technology for services to keep independence in different environment. Therefore, EdgeX Foundry and Azure IoT Edge choose to run as docker containers. As for Edgent, Edgent applications run on JVM.

\subsubsection{System characteristic}
System characteristics show the unique features of the system, which may help users to develop, deploy or monitor their edge applications. It will save lots of workload and time if making good use of these characteristics. EdgeX Foundry provides a common API to manage the devices, and this brings great convenience to deploying and monitoring edge applications in large scale. Azure IoT Edge provides powerful Azure services to accelerate the development of edge applications. Apache Edgent provides a series of functional APIs for data analytics, which lowers the difficulty and reduces the time on developing edge analytic applications. CORD and Akraino Edge Stack provide with multi-access edge services on edge cloud. We only need to keep connection with operator network, and we can apply for these services without the need to deploy an edge computing system on edge devices by ourselves. 

\subsubsection{Limitation}
This subsection discusses the limitation of the latest version of them to deploy edge applications. The lastest version of EdgeX Foundry has not provided a programmable interface in its architecture for developers to write their own applications. Although EdgeX allows us to add custom implementations, it demands more workload and time. As for Azure IoT Edge, though it is open-source and free, Azure services are chargeable as commercial software. For Apache Edgent, it is lightweight and it focuses on only data analytics. As for CORD and Akraino Edge Stack, these two systems demand stable network between data sources and the operators because the edge applications are running on the edge of operator network rather than local devices. 

\subsubsection{Scalability}
Increasing applications at edge make the network architecture more complex and the application management more difficult. Scalability is one major concern in edge computing. Among these edge computing systems, Azure IoT Edge, CORD and Akraino Edge Stack apply docker technology or virtual machine technology to support users to scale up or down their applications efficiently by adding or deleting module images. EdgeX Foundry is also a scalable platform that enables users to dynamically add or reduce microservices to adapt to the actual needs. But Apache Edgent is not scalable enough because every Edgent application is a single Java application and performance cannot be changed dynamically. 

\subsubsection{Mobility}
For EdgeX Foundry, Apache Edgent and Azure IoT Edge, once the applications are executed on some edge devices, they cannot be dynamically migrated to other devices. CORD and Akraino Edge Stack, deployed in the telecom infrastructure, support mobile edge services through mobile access network like 4G/5G. The mobility of these systems meet the need of cases such as unmanned cars and drones.

\subsubsection{Scenarios}
\begin{figure}[!tb]
\begin{center}
\includegraphics[width=0.48\textwidth]{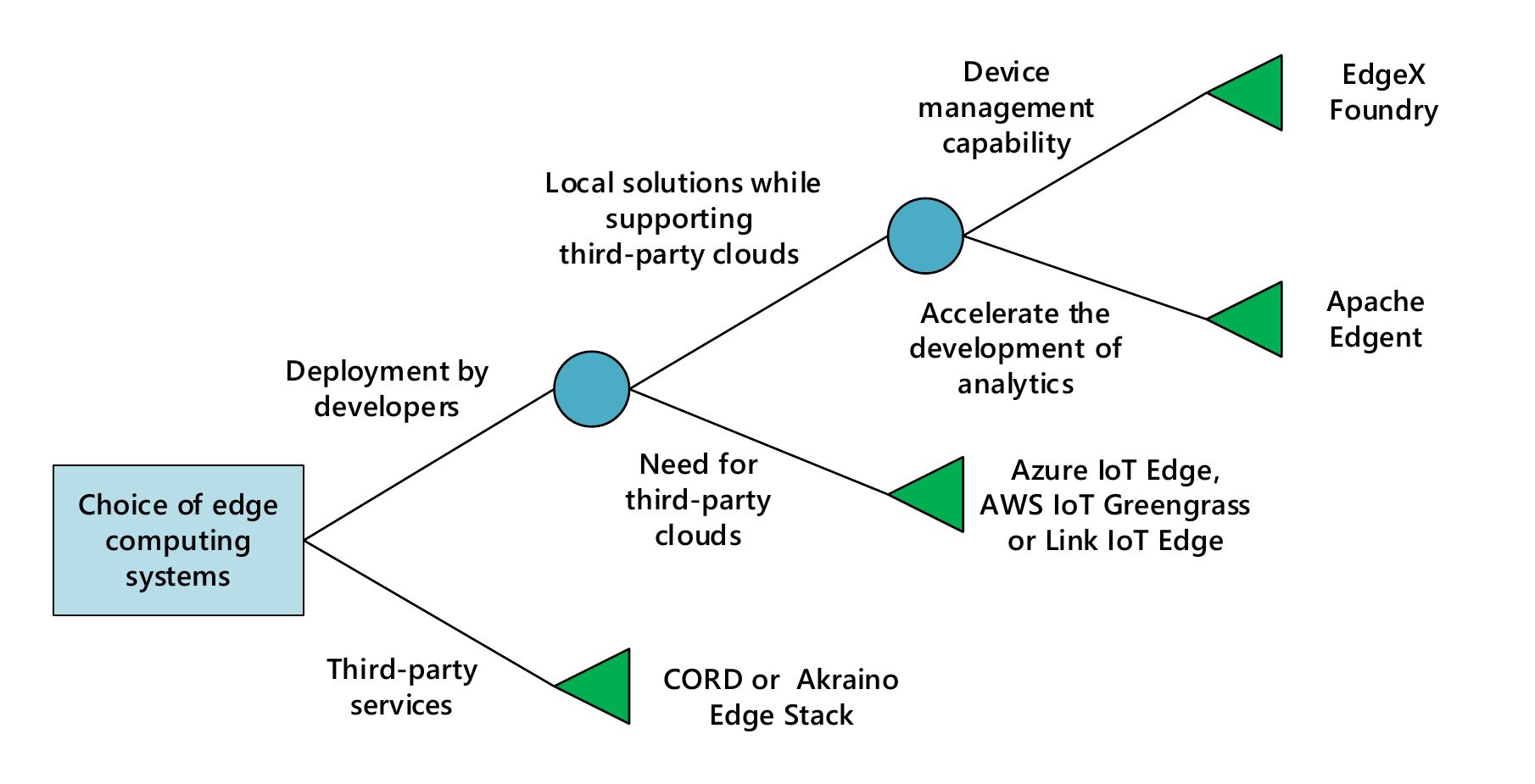}
\caption{The choice of open-source tools in different scenarios.}\label{fig:scenarios}
%\vspace{-0.20in}
\end{center}
\end{figure}

We discuss the choice of open-source tools from the perspective of the following three scenarios, as shown in Fig.~\ref{fig:scenarios}.

In the first scenario, suppose IoT edge applications are running on local area network (LAN), and the local enterprise system is regarded as back-end system with no need for third-party clouds. In this case, EdgeX Foundry or Apache Edgent are favorable because they enable users to build and control their own back-end system without being bound to any specific cloud platform. Furthermore, for the sake of managing and controlling edge devices on a large scale, EdgeX Foundry is a better choice for good device management capability. If considering data analysis, Apache Edgent is preferable to EdgeX Foundry. It provides a programming model to accelerate the development of edge analytic applications and a set of powerful APIs for data processing. 

In the second scenario, suppose the cloud services push to the edge of network to improve the quality. In this case, Auzre IoT Edge provides a convenient way for developers to migrate the applications from the cloud to the edge devices, and to leverage third-party high value services or functions when developing edge systems. Besides, AWS IoT Greengrass and Link IoT Edge, which are published by Amazon and Alibaba Cloud, are good choices as the competitive projects. More specifically, Azure IoT Edge provides Azure services such as Azure Functions, Azure Stream Analytics, and Azure Machine Learning. AWS IoT Greengrass can run AWS Lamda functions and machine learning models that are created, trained, and optimized in the cloud. Link IoT Edge provides Function Compute and other functions. Based on the application requirements, a suitable system could be chosen among them by taking account the functions they provide.

In the third scenario, suppose the users expect to directly leverage third-party services to deploy edge applications without any hardware or software system locally. In this case, edge systems such as CORD and Akraino Edge Stack are suitable. The users could choose one of them dependent on their application requirements. These systems are deployed by telecom operators. And telecom operators are also the providers of network services. Thus when there exist special network requirements of edge applications, these systems could satisfy the requirements. For example, edge computing applications on unmanned vehicles or drones need the support of wireless telecom networks (such as 4G/5G), in this case, a mobile edge computing service provided by these systems is a good choice.

In addition to the systems described above, there are other emerging open source projects. Device Management Edge, as part of Mbed IoT Platform published by ARM, is responsible for edge computing and provides the ability to access, manage and control Edge devices. KubeEdge, released by Huawei, provides edge nodes with native containerized application orchestration capabilities.

%Suppose we want to run edge applications on local area network, and use local enterprise system as back-end system with no need for third-party clouds. In this case, we can choose EdgeX Foundry or Apache Edgent. Further, suppose want to add various devices to the system and have good device management capabilities. We should choose EdgeX Foundry because it provides APIs to manage and control devices. If we focus on data analysis, a best approach would be to use Apache Edgent because it helps to accelerate the development of edge analytic applications.

%If we want to develop edge applications of artificial intelligence, Azure IoT Edge can reduce the difficulty of development by providing powerful Azure services like Azure Machine Learning, as well as commercial support. In addition, if we want to run mobile edge applications on drones or autonomous cars, we should choose edge cloud services with wireless access, and thus CORD or Akraino Edge Stack are the best choices.

\section{Enhancing Energy Efficiency of Edge Computing Systems}
\label{sec:energy-efficiency}

\begin{figure}[!tb]
\begin{center}
\includegraphics[width=0.32\textwidth]{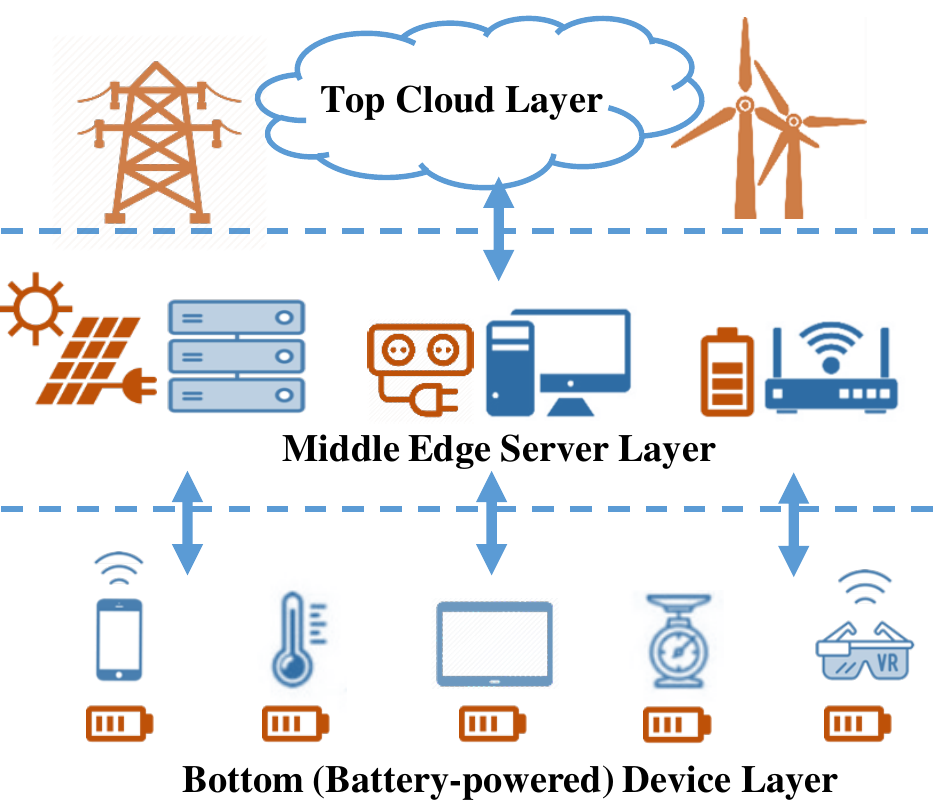}
\caption{The Three-layer Edge Computing Paradigm from a Power Source View.}\label{fig:three-layer-paradigm}
%\vspace{-0.20in}
\end{center}
\end{figure}

In the design of an edge computing system, energy consumption is always considered as one of the major concerns and evaluated as one of the key performance metrics. In this section, we review the energy efficiency enhancing mechanisms adopted by the state-of-the-art edge computing systems, from the view of \emph{top cloud layer}, \emph{middle edge server layer} and \emph{bottom device layer}, respectively (the three-layer paradigm is illustrated by Fig.~\ref{fig:three-layer-paradigm}).

\subsection{At the Top Cloud Layer}
For cloud computing, a centralized DC can comprise thousands of servers and thus consume enormous energy~\cite{mastelic2015cloud}. As an alternative to the cloud computing, does the edge/fog computing paradigm consume more or less energy? Different points of view have been given: some claim that decentralized data storage and processing supported by the edge computing architecture are more energy efficient~\cite{valancius2009greening,sarkar2016theoretical}, while some others show that such distributed content delivery may consume more energy than that of the centralized way~\cite{feldmann2010energy}.

The authors in~\cite{jalali2016fog} give a thorough energy analysis for applications running over the centralized DC (i.e., under cloud mode) and decentralized nano DCs (i.e., under fog mode), respectively. The results indicate that the fog mode may be with a higher energy efficiency, depending on several system design factors (e.g., type of application, type of access network, and ratio of active time), and those applications that generate and distribute a large amount of data in end-user premises result in best energy saving under the fog mode.

Note that the decentralized nano DCs or fog nodes in our context are different from the traditional CDN datacenters~\cite{tang2019tapping,tang2017rethinking,tang2018more}. They are also designed in a decentralized way but usually with much more powerful computing/communicating/storage capacities. 

\subsection{At the Middle Edge Server Layer}
At the middle layer of the edge computing paradigm, energy is also regarded as an important aspect, as the edge servers can be deployed in a domestic environment or powered by the battery (e.g., a desktop or a portable WiFi router, as shown in Fig.~\ref{fig:three-layer-paradigm}). Thus, to provide a higher availability, many power management techniques have been applied to limit the energy consumption of edge servers, while still ensuring their performances. We give a review of two major strategies used at the edge server layer in recent edge computing systems.

\subsubsection{Low-power system design and power management}
In~\cite{lewis2014tactical}, the tactical cloudlet is presented and its energy consumption when performing VM synthesis is evaluated particularly, under different cloudlet provisioning mechanisms. The results show that the largest amount of energy is consumed by i) VM synthesis due to the large payload size, and ii) on-demand VM provisioning due to the long application-ready time. Such results lead to the high energy efficiency policy: combining cached VM with cloudlet push for cloudlet provision.

A service-oriented architecture for fog/edge computing, Fog Data, is proposed and evaluated in~\cite{dubey2015fog}. It is implemented with an embedded computer system and performs data mining and data analytics on the raw data collection from the wearable sensors (in telehealth applications). With Fog Data, orders of magnitude data are reduced for transmission, thus leading to enormous energy saving. Furthermore, Fog Data is with a low power architecture design, and even consumes much less energy than that of a Raspberry Pi.

In~\cite{asnaghi2016dockercap}, a performance-aware orchestrator for Docker containers, named DockerCap, is developed to meet the power consumption constraints of the edge server (fog node). Following the observe-decide-act loop structure, DockerCap is able to manage container resources at run-time and provide soft-level power capping strategies. The experiments demonstrate that the obtained results with DockerCap is comparable to that from the power capping solution provided by the hardware (Intel RAPL).

An energy aware edge computer architecture is designed to be portable and usable in the fieldwork scenarios in~\cite{rausch2018portable}. Based on the architecture, a high-density cluster prototype is built using the compact general-purpose commodity hardware. Power management policies are implemented in the prototype to enable the real-time energy-awareness. Through various experiments, it shows that both the load balance strategies and cluster configurations have big impacts on the system energy consumption and responsiveness.

\subsubsection{Green-energy powered sustainable computing}
Dual energy sources are employed to support the running of a fog computing based system in~\cite{nan2017adaptive}, where solar power is utilized as the primary energy supply of the fog nodes. A comprehensive analytic framework is presented to minimize the long-term cost on energy consumption. Meanwhile, the framework also enables an energy-efficient data offloading (from fog nodes to the cloud) mechanism to help provide a high quality of service.

In~\cite{li2015towards}, a rack-scale green-energy powered edge infrastructure, InSURE (in-situ server system using renewable energy) is implemented for data pre-processing at the edge. InSURE can be powered by standalone (solar/wind) power and with batteries as the energy backup. Meanwhile, an energy buffering mechanism and a joint spatio-temporal power management scheme are applied, to enable efficient energy flow control from the power supply to the edge server.

\subsection{At the Bottom Device Layer}
As a well-recognized fact, the IoT devices in edge computing usually have strict energy constraints, e.g., limited battery life and energy storage. Thus, it remains a key challenge to power a great number (can up to tens of billions) of IoT devices at the edge, especially for those resource-intensive applications or services~\cite{mao2017survey}. We review the energy saving strategies adopted at the device layer of the edge computing diagram. Specifically, we go through three major approaches to achieving high energy efficiency in different edge/fog computing systems.

\subsubsection{Computation offloading to edge servers or cloud}
As a natural idea to solve the energy poverty problem, computation offloading from the IoT devices to the edge servers or cloud has been long investigated~\cite{rudenko1998saving,kemp2009eyedentify,cuervo2010maui}. It was also demonstrated that, for some particular applications or services, offloading tasks from IoT devices to more powerful ends can reduce the total energy consumption of the system, since the task execution time on powerful servers or cloud can be much shortened~\cite{ha2016system}. Although it increases the energy consumption of (wireless) data transmission, the tradeoff favors the offloading option as the computational demand increases~\cite{chun2011clonecloud}.

Having realized that the battery life is the primary bottleneck of handheld mobile devices, the authors in~\cite{cuervo2010maui} present MAUI, an architecture for mobile code offloading and remote execution. To reduce the energy consumption of the smartphone program, MAUI adopts a fine-grained program partitioning mechanism and minimizes the code changes required at the remote server or cloud. The ability of MAUI in energy reduction is validated by various experiments upon macro-benchmarks and micro-benchmarks. The results show that MAUI's energy saving for a resource-intensive mobile application is up to one order of magnitude, also with a significant performance improvement.

Like MAUI~\cite{cuervo2010maui} , the authors in~\cite{chun2011clonecloud} design and implement CloneCloud, a system that helps partition mobile application programs and performs strategically offloading for fast and elastic execution at the cloud end. As the major difference to MAUI, CloneCloud involves less programmer help during the whole process and only offloads particular program partitions on demand of execution, which further speeds up the program execution. Evaluation shows that CloneCloud can improve the energy efficiency of mobile applications (along with their execution efficiency) by 20 times. Similarly in~\cite{barbera2013offload}, by continuous updates of software clones in the cloud with a reasonable overhead, the offloading service can lead to energy reduction at the mobile end by a factor.
For computation intensive applications on resource constrained edge devices, their executions usually need to be offloaded to the cloud. To reduce the response latency of the image recognition application, Precog is presented which has been introduced in Sec.~\ref{subsec:Cachier_and_Precog}. With the on-device recognition caches, Precog much reduces the amount of images offloading to the edge server or cloud, by predicting and prefetching the future images to be recognized.

%For computation intensive applications on resource constrained edge devices, their executions usually need to be offloaded to the cloud. To reduce the response latency of the image recognition application, Precog is recently presented in~\cite{drolia2017precog}. With the on-device recognition caches, Precog much reduces the amount of images offloading to the edge server or cloud, by predicting and prefetching the future images to be recognized. As Precog also runs selective recognition tasks on the devices, the energy consumption is slightly more (approximately $5\%$ in average) than that of the device without Precog. The energy consumption from use of the camera and screen, which used to be the major cause of energy consumption in image recognition, does not show any increase in applying Precog.

\subsubsection{Collaborated devices control and resource management}
For energy saving of the massive devices at the edge, besides offloading their computational tasks to more powerful ends, there is also a great potential via sophisticated collaboration and cooperation among the devices themselves. Particularly, when the remote resources from the edge server or cloud are unavailable, it is critical and non-trivial to complete the edge tasks while without violating the energy constraint.

PCloud is presented in~\cite{jang2014personal} to enhance the capability of individual mobile devices at the edge. By seamless using available resources from the nearby devices, PCloud forms a ‘personal cloud’ to serve end users whenever the cloud resources are difficult to access, where device participation is guided in a privacy-preserving manner. The authors show that, by leveraging multiple nearby device resources, PCloud can much reduce the task execution time as well as energy consumption. For example, in the case study of ‘neighborhood watch with face recognition’, the results show a $74\%$ reduction in energy consumption on a PCloud vs. on a single edge device. Similar to PCloud, the concept of mobile device cloud (MDC) is proposed in~\cite{fahim2013making}, where computational offloading is also adopted among the mobile devices. It shows that the energy efficiency (gain) is increased by $26\%$ via offloading in MDC. The authors of~\cite{liu2014adaptive} propose an adaptive method to dynamically discovery available nearby resource in heterogeneous networks, and perform automatic transformation between centralized and flooding strategies to save energy.

As current LTE standard is not optimized to support a large simultaneous access of IoT devices, the authors in~\cite{abdelwahab2016replisom} propose an improved memory replication architecture and protocol, REPLISON, for computation offloading of the massive IoT devices at the edge. REPLISON improves the memory replication performance through an LTE-optimized protocol, where Device-to-Device (D2D) communication is applied as an important supporting technology (to pull memory replicas from IoT devices). The total energy consumption of REPLISON is generally worse than the conventional LTE scenario, as it needs more active devices. However, the energy consumed per device during a single replicate transmission is much less. With further evaluation results, it shows that REPLISOM has an energy advantage over the conventional LTE scenarios as long as the size of replica is sufficiently small.

For IoT devices distributed at the edge, the authors in~\cite{teemu2018energy} leverage the software agents running on the IoT devices to establish an integrated multi-agent system (MAS). By sharing data and information among the mobile agents, edge devices are able to collaborate with each other and improve the system energy efficiency in executing distributed opportunistic applications. Upon the experimental platform with 100 sensor nodes and 20 smartphones as edge devices, the authors show the great potential of data transmission reduction with MAS. This leads to a significant energy saving, from $15\%$ to $66\%$, under different edge computing scenarios. As another work applying data reduction for energy saving, CAROMM~\cite{jayaraman2014cardap} employs a change detect technique (LWC algorithm) to control the data transmission of IoT devices, while maintaining the data accuracy.

\section{Deep Learning Optimization at the Edge}
\label{sec:deep-learning-opt}

In the past decades, we have witnessed the burgeoning of machine learning, especially deep learning based applications which have changed human being's life. With complex structures of hierarchical layers to capture features from raw data, deep learning models have shown outstanding performances in those novel applications, such as machine translation, object detection, and smart question and answer systems.

Traditionally, most deep learning based applications are deployed on a remote cloud center, and many systems and tools are designed to run deep learning models efficiently on the cloud. Recently, with the rapid development of edge computing, the deep learning functions are being offloaded to the edge. Thus it calls for new techniques to support the deep learning models at the edge. This section classifies these technologies into three categories: systems and toolkits, deep learning packages, and hardware.

\subsection{Systems and Toolkits}
Building systems to support deep learning at the edge is currently a hot topic for both industry and academy. There are several challenges when offloading state-of-the-art AI techniques on the edge directly, including computing power limitation, data sharing and collaborating, and mismatch between edge platform and AI algorithms. To address these challenges, OpenEI is proposed as an Open Framework for Edge Intelligence~\cite{zhang2019openei}. OpenEI is a lightweight software platform to equip edges with intelligent processing and data sharing capability. OpenEI consists of three components: a package manager to execute the real-time deep learning task and train the model locally, a model selector to select the most suitable model for different edge hardware, and a library including a RESTFul API for data sharing. The goal of OpenEI is that any edge hardware will has the intelligent capability after deploying it. 

In the industry, some top-leading tech-giants have published several projects to move the deep learning functions from the cloud to the edge. Except Microsoft published Azure IoT Edge which have been introduced in Sec.~\ref{subsec:AzureIoT}, Amazon and Google also build their services to support deep learning on the edge.
Table~\ref{table: Comparison of Deep learning Systems on Edge} summarizes the features of the systems which will be discussed below.

Amazon Web Services (AWS) has published IoT Greengrass ML Inference~\cite{Amazonedge} after IoT Greengrass. AWS IoT Greengrass ML Inference is a software to support machine learning inferences on local devices. With AWS IoT Greengrass ML Inference, connected IoT devices can run AWS Lambda functions and have the flexibility to execute predictions based on those deep learning models created, trained, and optimized in the cloud. AWS IoT Greengrass consists of three software distributions: AWS IoT Greengrass Core, AWS IoT Device SDK, and AWS IoT Greengrass SDK. Greengrass is flexible for users as it includes a pre-built TensorFlow, Apache MXNet, and Chainer package, and it can also work with Caffe2 and Microsoft Cognitive Toolkit. 

Cloud IoT Edge~\cite{Cloud_IoT_Edge} extends Google Cloud's data processing and machine learning to edge devices by taking advantages of Google AI products, such TensorFlow Lite and Edge TPU. Cloud IoT Edge can either run on Android or Linux-based operating systems. It is made up of three components: Edge Connect ensures the connection to the cloud and the updates of software and firmware, Edge ML runs ML inference by TensorFlow Lite, and Edge TPU specific designed to run TensorFlow Lite ML models. Cloud IoT Edge can satisfy the real-time requirement for the mission-critical IoT applications, as it can take advantages of Google AI products (such as TensorFlow Lite and Edge TPU) and optimize the performance collaboratively.

\begin{table*}[!htp]
	\caption{Comparison of Deep learning Systems on Edge}
	\label{table: Comparison of Deep learning Systems on Edge}
	\begin{tabular}{ l l l l }
		\hline
		\textbf{Features} & \textbf{AWS IoT Greengrass} & \textbf{Azure IoT Edge} & \textbf{Cloud IoT Edge} \\ \hline
		Developer & Amazon & Microsoft & Google \\ \hline
		Components & \begin{tabular}[c]{@{}l@{}}IoT Greengrass Core, IoT Device SDK,  \\ IoT Greengrass SDK\end{tabular} & \begin{tabular}[c]{@{}l@{}}IoT Edge modules, IoT Edge runtime,  \\ Cloud-based interface\end{tabular} & \begin{tabular}[c]{@{}l@{}}Edge Connect, Edge ML, Edge TPU\end{tabular} \\ \hline
		OS & Linux, macOS, Windows & Windows, Linux, macOS & Linux, macOS, Windows, Android \\ \hline
		Target device & Multiple platforms (GPU-based, Raspberry Pi) & Multiple platforms & TPU\\ \hline
		Characteristic & Flexible & Windows friendly & Real-time \\ \hline
	\end{tabular}
\end{table*}

\subsection{Deep Learning Packages}
Many deep learning packages have been widely used to deliver the deep learning algorithms and deployed on the cloud data centers, including TensorFlow~\cite{abadi2016tensorflow}, Caffe~\cite{jia2014caffe}, PyTorch~\cite{PyTorch}, and MXNet~\cite{chen2015mxnet}. Due to the limitations of computing resources at the edge, the packages designed for the cloud are not suitable for edge devices. Thus, to support data processing with deep learning models at the edge, several edge-based deep learning frameworks and tools have been released. In this section, we introduce TensorFlow Lite, Caffe2, PyTorch, MXNet, CoreML~\cite{CoreML}, and TensorRT~\cite{TensorRT}, whose features are summarized in Tables \ref{table: Comparison of Deep Learning Frameworks on Edge}.

TensorFlow Lite~\cite{TensorFlow-Lite} is TensorFlow's lightweight solution which is designed for mobile and edge devices. TensorFlow is developed by Google in 2016 and becomes one of the most widely used deep learning frameworks in cloud data centers. To enable low-latency inference of on-device deep learning models, TensorFlow Lite leverages many optimization techniques, including optimizing the kernels for mobile apps, pre-fused activations, and quantized kernels that allow smaller and faster (fixed-point math) models. 

Facebook published Caffe2~\cite{Caffe2} as a lightweight, modular, and scalable framework for deep learning in 2017. Caffe2 is a new version of Caffe which is first developed by UC Berkeley AI Research (BAIR) and community contributors. Caffe2 provides an easy and straightforward way to play with the deep learning and leverage community contributions of new models and algorithms. Comparing with the original Caffe framework, Caffe2 merges many new computation patterns, including distributed computation, mobile, reduced precision computation, and more non-vision use cases. Caffe2 supports multiple platforms which enable developers to use the power of GPUs in the cloud or at the edge with cross-platform libraries.

PyTorch~\cite{PyTorch} is published by Facebook. It is a python package that provides two high-level features: tensor computation with strong GPU acceleration and deep Neural Networks built on a tape-based auto-grad system. Maintained by the same company (Facebook), PyTorch and Caffe2 have their own advantages. PyTorch is geared toward research, experimentation and trying out exotic neural networks, while caffe2 supports more industrial-strength applications with a heavy focus on the mobile. In 2018, Caffe2 and PyTorch projects merged into a new one named PyTorch 1.0, which would combine the user experience of the PyTorch frontend with scaling, deployment and embedding capabilities of the Caffe2 backend.

MXNet~\cite{chen2015mxnet} is a flexible and efficient library for deep learning. It was initially developed by the University of Washington and Carnegie Mellon University, to support CNN and long short-term memory networks (LSTM). In 2017, Amazon announced MXNet as its choice of deep learning framework. MXNet places a special emphasis on speeding up the development and deployment of large-scale deep neural networks. It is designed to support multiple different platforms (either cloud platforms or the edge ones) and can execute training and inference tasks.  Furthermore, other than the Windows, Linux, and OSX operating systems based devices, it also supports the Ubuntu Arch64 and Raspbian ARM based operating systems.

CoreML~\cite{CoreML} is a deep learning framework optimized for on-device performance at memory footprint and power consumption. Published by Apple, users can integrate the trained machine learning model into Apple products, such as Siri, Camera, and QuickType. CoreML supports not only deep learning models, but also some standard models such as tree ensembles, SVMs, and generalized linear models. Built on top of low level technologies, CoreML aims to make full use of the CPU and GPU capability and ensure the performance and efficiency of data processing.

The platform of TensorRT~\cite{TensorRT} acts as a deep learning inference to run the models trained by TensorFlow, Caffe, and other frameworks. Developed by NVIDIA company, it is designed to reduce the latency and increase the throughput when executing the inference task on NVIDIA GPU. To achieve computing acceleration, TensorRT leverages several techniques, including weight and activation precision calibration, layer and tensor fusion, kernel auto-tuning, dynamic tensor memory, and multi-stream execution.

Considering the different performance of the packages and the diversity of the edge hardware, it is challenging to choose a suitable package to build edge computing systems. To evaluate the deep learning frameworks at the edge and provide a reference to select appropriate combinations of package and edge hardware, pCAMP~\cite{pCAMP} is proposed. It compares the packages' performances (w.r.t. the latency, memory footprint, and energy) resulting from five edge devices and observes that no framework could win over all the others at all aspects.
It indicates that there is much room to improve the frameworks at the edge. Currently, developing a lightweight, efficient and high-scalability framework to support diverse deep learning modes at the edge cannot be more important and urgent.

In addition to these single-device based frameworks, more researchers focus on distributed deep learning models over the cloud and edge. DDNN~\cite{teerapittayanon2017distributed} is a distributed deep neural network architecture across cloud, edge, and edge devices. DDNN maps the sections of a deep neural network onto different computing devices, to minimize communication and resource usage for devices and maximize usefulness of features extracted from the cloud. 

Neurosurgeon~\cite{kang2017neurosurgeon} is a lightweight scheduler which can automatically partition DNN computation between mobile devices and datacenters at the granularity of neural network layers. By effectively leveraging the resources in the cloud and at the edge, Neurosurgeon achieves low computing latency, low energy consumption, and high traffic throughput. 
% The experimental results show that Neurosurgeon improves end-to-end latency by 3.1X, reduces mobile energy consumption by 59.5\%, and improves datacenter throughput by 1.5X on average.

\begin{table*}[!htp]
	\centering
	\caption{Comparison of Deep Learning Packages on Edge}
	\label{table: Comparison of Deep Learning Frameworks on Edge}
	\begin{tabular}{ l l l l l l l }
		\hline
		\textbf{Features} & \textbf{TensorFlow Lite} & \textbf{Caffe2} & \textbf{PyTorch} & \textbf{MXNet} & \textbf{CoreML} & \textbf{TensorRT} \\ \hline
		Developer & Google & Facebook & Facebook & DMLC, Amazon & Apple & NVIDIA \\ \hline
		\begin{tabular}[c]{@{}l@{}}Open Source \\ License\end{tabular} & Apache-2.0 & Apache-2.0 & BSD & Apache-2.0 & Not open source & Not open source \\ \hline
		Task & Inference & Training, Inference & Training, Inference & Training, Inference & Inference & Inference \\ \hline
		Target Device & \begin{tabular}[c]{@{}l@{}}Mobile and \\ embedded device\end{tabular} & Multiple platform & Multiple platform & Multiple platform & Apple devices & NVIDIA GPU \\ \hline
		Characteristic & Latency & \begin{tabular}[c]{@{}l@{}}Lightweight, modular, \\ and scalable\end{tabular} & Research & \begin{tabular}[c]{@{}l@{}}Large-scale \\ deep neural networks\end{tabular} & \begin{tabular}[c]{@{}l@{}}Memory footprint and \\ power consumption.\end{tabular} & \begin{tabular}[c]{@{}l@{}}Latency and\\  throughput\end{tabular} \\ \hline
	\end{tabular}
\end{table*}

\subsection{Hardware System}
The hardware designed specifically for deep learning can strongly support edge computing. Thus, we further review relevant hardware systems and classify them into three categories: FPGA-based hardware, GPU-based hardware, and ASIC.

\subsubsection{FPGA-based Hardware}

A field-programmable gate array (FPGA) is an integrated circuit and can be configured by the customer or designer after manufacturing. FPGA based accelerators can achieve high performance computing with low energy, high parallelism, high flexibility, and high security~\cite{wang2018survey}.

\cite{zhang2015optimizing} implements a CNN accelerator on a VC707 FPGA board. The accelerator focuses on solving the problem that the computation throughput does not match the memory bandwidth well. By quantitatively analyzing the two factors using various optimization techniques, the authors provide a solution with better performance and lower FPGA resource requirement, and their solution achieves a peak performance of $61.62$ GOPS under a $100MHz$ working frequency. 

Following the above work, Qiu et al.~\cite{qiu2016going} propose a CNN accelerator designed upon the embedded FPGA, Xilinx Zynq ZC706, for large-scale image classification. It presents an in-depth analysis of state-of-the-art CNN models and shows that Convolutional layers are computational-centric and Fully-Connected layers are memory-centric. The average performances of the CNN accelerator at convolutional layers and the full CNN are $187.8$ GOPS and $137.0$ GOPS under a $150MHz$ working frequency, respectively, which outperform previous approaches significantly.

An efficient speech recognition engine (ESE) is designed to speed up the predictions and save energy when applying the deep learning model of LSTM. ESE is implemented in a Xilinx XCKU060 FPGA opearting at $200MHz$. For the sparse LSTM network, it can achieve 282GOPS, corresponding to a $2.52$ TOPS on the dense LSTM network. Besides, energy efficiency improvements of $40x$ and $11.5x$ are achieved, respectively, compared with the CPU and GPU based solution.

\subsubsection{GPU-based hardware}

GPU can execute parallel programs at a much higher speed than CPU, which makes it fit for the computational paradigm of deep learning algorithms. Thus, to run deep learning models at the edge, building the hardware platform with GPU is a must choice. Specifically, NVIDIA Jetson TX2 and DRIVE PX2 are two representative GPU-based hardware platforms for deep learning.

NVIDIA Jetson TX2~\cite{JetsonTX2} is an embedded AI computing device which is designed to achieve low latency and high power-efficient. It is built upon an NVIDIA Pascal GPU with 256 CUDA cores, an HMP Dual Denver CPU and a Qualcomm ARM CPU. It is loaded with 8GB of memory and 59.7GB/s of memory bandwidth and the power is about 7.5 watts. The GPU is used to execute the deep learning task and CPUs are used to maintain general tasks. It also supports the NVIDIA Jetpack SDK which includes libraries for deep learning, computer vision, GPU computing, and multimedia processing.

NVIDIA DRIVE PX~\cite{NVIDIA_DRIVE_PX} is designed as the AI supercomputer for autonomous driving. The architecture is available in a variety of configurations, from the mobile processor operating at 10 watts to a multi-chip AI processors delivering 320 TOPS. It can fuse data from multiple cameras, as well as lidar, radar, and ultrasonic sensors.

\subsubsection{ASIC}
Application-Specific Integrated Circuit (ASIC) is the integrated circuit which supports customized design for a particular application, rather than the general-purpose use. ASIC is suitable for the edge scenario as it usually has a smaller size, lower power consumption, higher performance, and higher security than many other circuits. Researchers and developers design ASIC to meet the computing pattern of deep learning.

DianNao family~\cite{chen2016diannao} is a series of hardware accelerators designed for deep learning models, including DianNao, DaDianNao, ShiDianNao, and PuDianNao. They investigate the accelerator architecture to minimize memory transfers as efficiently as possible. Different from other accelerators in DianNao family, ShiDianNao~\cite{du2015shidiannao} focuses on image applications in embedded systems which are widely used in edge computing scenarios. For the CNN-based deep learning models, it provides a computing speed $30x$ faster than that of NVIDIA K20M GPU, with an body area of $4.86 mm^2$ and a power of $320 mW$.

Edge TPU~\cite{EdgeTPU} is Google's purpose-built ASIC for edge computing. It augments Google's Cloud TPU and Cloud IoT to provide an end-to-end infrastructure and facilitates the deployment of customers' AI-based solutions. In addition, Edge TPU can combine the custom hardware, open software, and state-of-the-art AI algorithms to achieve high performance with a small physical area and low power consumption.

\section{Key Design Issues}
\label{sec:key-design-issues}

The edge computing system manages various resources along the path from the cloud center to end devices, shielding the complexity and diversity of hardware and helping developers quickly design and deploy novel applications. To fully leverages the advantages, we discuss the following key issues which need to be paid attention when analyzing and designing a new edge computing system.

\subsubsection{Mobility Support}
Mobility support has two aspects: user mobility and resource mobility. User mobility refers to how to automatically migrate the current program state and necessary data when the user moves from one edge node coverage domain to another so that the service handover is seamlessly. Currently, Cloudlet and CloudPath provide service migration through terminating/finishing existing task and starting a new VM/instance in the target edge node. Fine-grain migration and target edge node prediction are not supported. Resource mobility refers to i) how to dynamically discover and manage available resources, including both the long-term resources and short-term ones, and ii) when some edge devices are damaged how the system can be resumed as soon as possible with replacements. For example, PCloud and FocusStack supports edge devices dynamic join and leave with no task running. Intelligent dynamic resource management is still required for edge computing systems.

\subsubsection{Multi-user Fairness}
For edge devices with limited resources, how to ensure the fairness of multi-user usage, especially for the shared resources and rare resources. For example, a smartphone made up of various sensors and computing resources can act as an edge node to serve multiple users. However, as the smartphone has limited battery life, it is a problem to fairly allocate resources when receiving many requests. The resource competition is more intense, the requests are coming from different irrelevant tasks, it is hard to decide who should use the resource with only local information. Besides, the unfairness can be used as an attack strategy to make the critical task resource hungry, which may lead to main task failure. The existing edge computing systems do not pay much attention to the multi-user fairness, the basic solution (bottom-line) is to provide resource isolation, users only get what the edge node promised when it accepts the request. More fairness strategies require system support, like related task status updates, etc.

\subsubsection{Privacy Protection}
Unlike cloud computing, edge devices can be privately owned, such as gateway devices for smart home systems. When other users use such edge devices, obtain their data, and even take control of them, how to ensure the owner's privacy and guest users' data privacy is important. AirBox is a lightweight flexible edge function system, which leverages hardware security mechanism (e.g. Intel SGX) to enhance system security. Other system have not pay much attention to privacy and security. Existing cloud security approaches can be applied with the consideration of the resource limitation. Enhancing resource isolation, setting up privilege management and access control policies can be potential directions to solve the privacy problem. 

\subsubsection{Developer Friendliness}
The system ultimately provides hardware interaction and basic services for upper-level applications. How to design interactive APIs, program deployment module, resource application and revocation, etc., are the key factors for the system to be widely used. Therefore, to design an edge computing system, we should think from an application developer's perspective. Specifically, to provide effective development and deployment services is a good idea to help improve the ecosystem of the edge computing system. For instances, EdgeX Foundry and Apache Edgent provide APIs to manage devices and analyze data respectively. ParaDrop supports monitoring devices and application status through developer APIs, and provides a web UI for users to manage/configure gateway as well as deploy applications to devices. 

\subsubsection{Multi-domain Management and Cooperation}
Edge computing involves multiple types of resources, each of which may belong to a different owner. For examples, the smart home gateways and sensors of the house owner, networks resources and base stations from the Internet service providers (ISP), traffic cameras owned by the government. How to access these resources and organize them according to the requirements of application and services, especially in emergency situations, is a problem that the edge computing system needs to consider. Existing researches assume we have the permission to use various devices belongs to different owners. In the initial stage, systems focus on the functionality perspective rather than implementation issues like price model. With the developing of edge computing, the real deployment issues should be solved before we enjoy all the benefits.

\subsubsection{Cost Model}
In cloud computing, the corresponding virtual machine can be allocated based on the resource requested by the user, and the cost model can be given according to the resource usage. In edge computing, an application may use resources from different owners. Thus, how to measure resource usage, calculate overall overhead, and give an appropriate pricing model are crucial problems when deploying an edge computing system. Generally, edge computing systems focus on how to satisfy the resource requirements of various services and applications, some of them pay more attention to the energy consumption due to the nature of mobile nodes, more complex overhead are not considered.

\subsubsection{Compatibility}
Currently, specialized edge computing applications are still quite few. For examples, ParaDrop applications require additional XML configuration file to specify the resource usage requirements, SpanEdge needs developers to divide/configure tasks into local tasks and global tasks. Common applications are not directly supported to run in edge systems. How to automatically and transparently convert existing programs to the edge version and conveniently leverages the advantages of edge computing are still open problems. The compatibility should be considered in the edge system design. Specifically, how to adapt traditional applications to the new architecture and realize the basic functions so that they can run successfully, deserving further exploration and development.

\section{Conclusions}\label{sec:conclusion}
Edge computing is a new paradigm that jointly migrates the capability of networking, computation, and storage from the remote cloud to the user sides. Under the context of IoT and 5G, the vision of edge computing is promising in providing smarter services and applications with better user experiences. The recently proposed systems and tools on edge computing generally reduce the data processing and transmission overhead, and improve the efficiency and efficacy of mobile data analytics. Additionally, the integration of edge computing and deep learning techniques further fosters the research on edge-based intelligence services. This article introduced the representative systems and open source projects for edge computing, presented several energy efficiency enhancing strategies for performance consideration and technologies for deploying deep learning models at the edge, and suggested a few research directions during system design and analysis.

\nop{
\section*{Acknowledgment}
This work is partially supported by the National Key Research and Development Program of China (2016YFB1000302) and National Natural Science Foundation of China (61433019). Fang Liu and
Zhiping Cai are the corresponding authors.
}

\bibliographystyle{IEEEtran}
\bibliography{reference}

\end{document}